\documentclass[preprint,12pt]{elsarticle}
\usepackage{dsfont,amssymb,amsmath,graphics,graphicx,fancyhdr,hyperref, float,verbatim,
ifthen,ifpdf,algorithmic,multirow,array,color, enumerate,theorem}
\usepackage[english]{babel}
\usepackage[perpage]{footmisc}
\usepackage[ruled]{algorithm}
\usepackage[subnum]{cases}

\renewcommand{\parallel}{|\!|}

\newcommand{\bh}{\mathbf{h}}

\newcommand{\bt}{\mathbf{t}}

\newcommand{\bT}{\mathbf{T}}

\newcommand{\bw}{\mathbf{w}}

\newcommand{\bx}{\mathbf{x}}

\newcommand{\bz}{\mathbf{z}}

\newcommand{\bsw}{\boldsymbol{w}}

\newcommand{\cL}{\mathcal{L}}

\newcommand{\bsmu}{\boldsymbol{\mu}}

\newcommand{\bsbeta}{\boldsymbol{\beta}}

\newcommand{\bsSigma}{\boldsymbol{\Sigma}}

\newcommand{\bsPsi}{\boldsymbol{\Psi}}

\newcommand{\Identity}{\textbf{I}}

\newcommand{\E}{\mathbb{E}}

\newcommand{\R}{\mathbb{R}}
	
\newcommand{\N}{\mathcal{N}}

\graphicspath{{Figs/}}

\journal{Neurocomputing Elsevier}

\begin{document}

\begin{frontmatter}
 
\title{Model-based functional mixture discriminant analysis with hidden process regression for curve classification} 

\author[LSIS-USTV,LSIS-AMU]{F. Chamroukhi \corref{cor1}} \ead{faicel.chamroukhi@univ-tln.fr}
\cortext[cor1]{Corresponding author: Faicel Chamroukhi\\ Universit\'e de Toulon, LSIS, UMR CNRS 7296 \\  B\^atiment R, BP 20132 - 83957 La Garde Cedex, France \\Tel: +33(0) 4 94 14 20 06\\Fax: +33(0) 4 94 14 28 97 }
\author[LSIS-USTV,LSIS-AMU,IUF]{H. Glotin} 

\address[LSIS-USTV]{Universit\'e de Toulon, CNRS, LSIS, UMR 7296, 83957 La Garde, France} 
\address[LSIS-AMU]{Aix Marseille Universit\'e, CNRS, ENSAM, LSIS, UMR 7296, 13397 Marseille, France}
\address[IUF]{Institut Universitaire de France, iuf.amue.fr}
\author[Ifsttar]{A. Sam\'e}
\address[Ifsttar]{UPE, IFSTTAR, GRETTIA, France}

\begin{abstract} 
In this paper, we study the modeling and the classification of functional data  presenting regime changes over time. We propose a new model-based functional mixture discriminant analysis approach based on a specific hidden process regression model that governs the regime changes over time. Our approach is particularly adapted to handle the problem of complex-shaped classes of curves, where each class is potentially composed of several sub-classes, and to deal with the regime changes within each homogeneous sub-class. The proposed model explicitly integrates the heterogeneity of each class of curves via a mixture model formulation, and the regime changes within each sub-class through a hidden logistic process. Each class of complex-shaped curves is modeled by a finite number of homogeneous clusters,  each of them being decomposed into several regimes. The model parameters of each class are learned by maximizing the observed-data log-likelihood  by using a dedicated expectation-maximization (EM) algorithm. Comparisons are performed with alternative curve classification approaches, including functional linear discriminant analysis and functional mixture discriminant analysis with polynomial regression mixtures and spline regression mixtures. Results obtained on simulated data and real data show that the proposed approach outperforms the alternative approaches in terms of discrimination,  and significantly improves the curves approximation.
\end{abstract}


\end{frontmatter}

\section{Introduction}

Most statistical analyses involve vectorial data when the observations are finite dimensional vectors. However, in many application domains, such as diagnosis of complex systems \citep{chamroukhi_et_al_neurocomputing2010,chamroukhi_adac_2011}, electrical engineering \citep{hebrail_etal_Neurocomputing2010}, speech recognition (e.g. the phoneme data studied in \citep{Delaigle2012}), radar waveform \citep{DaboNiang2007}, etc, the data are functions (.i.e curves) rather than finite dimensional vectors. The paradigm of analysing such data is known as Functional Data Analysis (FDA) \citep{ramsayandsilvermanFDA2005}. The statistical approaches for FDA concern the analysis of data for which the individuals are entire functions or curves rather than finite dimensional vectors. The goals of FDA include data representation for further analysis,  data visualization, exploratory analysis by performing clustering or projections, regression, classification, etc. 
 Additional background on FDA, examples and analysis techniques can be found in \cite{ramsayandsilvermanFDA2005}. 
 The analysis task leads in general to learning a statistical model namely in a supervised context for classification, regression, or in an unsupervised way for a clustering or a segmentation task \citep{gaffneyANDsmythNIPS2004, chamroukhi_et_al_neurocomputing2010, chamroukhi_adac_2011, GaffneyThesis, chamroukhi_PhD_2010, Gui_FMDA, garetjamesANDtrevorhastieJRSS2001, Delaigle2012}, etc. 
The challenge is therefore to build adapted models to be learned from such data living in a very high or an infinite dimensional space. 
In this paper, we consider the problem of supervised functional data classification (discrimination) where the observations are temporal curves presenting several regime changes over time. 
However, while the global task is supervised, as we shall present it later, this global discrimination task includes two unsupervised tasks. The first one is to automatically cluster possibly dispersed classes into several homogeneous clusters (i.e., sub-classes), and the second one is to automatically determine the temporal regimes of each sub-class which can be seen as a temporal segmentation task.

 Indeed, concerning the first point of class dispersion, that is the need of sub-classses, in many areas of application of classification, a class itself may be composed of several unknown (unobserved) sub-classes. The learning has therefore to be treated in an unsupervised way within each class, since no labels of sub-groups are available. For example, in handwritten digit recognition, there are several characteristic ways to write a digit, and therefore a creation of several sub-classes within the class of digit itself, which may be modeled using a mixture density as in \cite{hastieANDtibshiraniMDA}. In complex systems diagnosis application,
where we have to decide between two classes :  without defect/wit defect, one would have only the class labels as either with or without defect, however no labels according to how a defect would happen, namely the type of defect, the degree of defect, means minor, critical, etc. Thus, providing an automatic tool that decomposes the class into sub-classes would be very helpful in making accurate decisions as well as for well interpretation.  
Another example is the one of gene function classification based on time course gene
expression data. As stated in \cite{Gui_FMDA} when considering the complexity of the gene functions, one functional class may includes genes which involve one
or more biological profiles. Describing each class as a combination of sub-classes, which unfortunately are very often unknown, is necessary to provide realistic description, rather than providing a rough representation.

We mainly focus on generative approaches, in particular latent variable models, which are dedicated to explain the  underlying processes generating the data. As it should be explicitly described later, this can be achieved by explicitly integrating the problem of class dispersion and the one of regimes changes over time into a two-level latent data model. 
The generative approaches for functional data related to this work are essentially based on regression analysis, including polynomial 
regression, splines and B-splines \citep{GaffneyThesis, chamroukhi_PhD_2010, Gui_FMDA, garetjamesANDtrevorhastieJRSS2001}, or also generative polynomial piecewise regression as in \cite{chamroukhi_PhD_2010, chamroukhi_et_al_neurocomputing2010}. Non-parametric statistical approaches have also been proposed for functional data discrimination as in \cite{FerratyV03,Delaigle2012} and clustering as in \cite{Delaigle2012}.
Another possible curve modeling can be the Gaussian processes approach \citep{RassmussenAndwillams_2006} which is a non parametric approach that has been used in functional data analysis \citep{ShiMT05_Hierarchical_GPR, Shi_etal_GPR, ShiGPR_Book2011}, one can site in particular recent Gaussian Processes for functional regression \citep{ShiGPR_Book2011}. While they are used as a non parametric approach, inference in such models requires performing MCMC sampling and direct implementation is computationally expensive. In this paper, we focused on parametric approaches where the computation of conditional expectations is analytic. The model parameters can be further used for summarizing a set of curves into a parameter vector. This is useful for example for a feature extraction prospective \citep{chamroukhiIJCNN2009}.  
Furthermore, while Gaussian process approaches are well adapted to approximate and to cluster non-linear functions or curves as in Hierarchical Gaussian process mixtures for regression \citep{ShiGPR_Book2011, ShiMT05_Hierarchical_GPR}, the problem of regime changes within each set of curves is still not taken into account in such approaches; only a non-linear approximation is provided, without segmentation. 
In this paper, we propose a new generative approach for modeling classes of complex-shaped curves where each class is itself composed of unknown homogeneous sub-classes. In addition, the model is particularly dedicated to address the problem when each homogeneous sub-class presents regime changes over time. 
Here we extend the functional discriminant analysis approach presented in \cite{chamroukhi_et_al_neurocomputing2010}, which relates modeling each class of curves presenting regime changes with a single mean curve, to a mixture formulation which leads to a functional mixture-model based discriminant analysis. More specifically, this approach uses a mixture of regression models with hidden logistic processes (RHLP) \citep{chamroukhi_PhD_2010, chamroukhi_adac_2011} for each class of functional data, and derives a functional mixture discriminant analysis framework for functional data classification. The resulting discrimination approach is therefore a model-based functional discriminant analysis in which learning the parameters of each class  of curves is achieved through an unsupervised estimation of a mixture of RHLP (MixRHLP) models. A first idea of this approach was presented in \cite{chamroukhi_PhD_2010, chamroukhi_esann2012}. %

In the next section we give a brief background on discriminant analysis approaches for functional data  classification  including functional linear and mixture discriminant analysis, 
and then we present the proposed model-based functional mixture discriminant analysis with hidden process regression for curve classification, which we will abbreviate as  FMDA-MixRHLP, and the corresponding parameter estimation procedure using a dedicated expectation-maximization (EM) algorithm. Then, we will present the model selection using the Bayesian Information Criterion (BIC) \citep{BIC}. 

In the following we denote by $((\bx_1,y_1),\ldots,(\bx_n,y_n))$ a given labeled training set of curves issued from $G$ classes where $y_i \in  \{1,\ldots,G \}$ is the class label of the $i$th curve $\bx_i$. We assume that $\bx_i$ consists of $m$ observations $(x_{i1},\ldots,x_{im})$, regularly observed at the time points $(t_1,\ldots,t_m)$ with $t_1<\ldots<t_m$. 
 
\section{Background on Functional Discriminant Analysis}
In this section, we give a background on generative discriminant analysis approaches for functional data classification based on functional regression.
Functional discriminant analysis approaches extend discriminant analysis approaches for vectorial data to functional data or curves.  
The functional discriminant analysis principle is as follows. Assume we have  a labeled training set of curves and the classes' parameter vectors $(\bsPsi_1,\ldots,\bsPsi_G)$ where $\bsPsi_g$ is the parameter vector of the density of class $g$ $(g=1,\ldots,G)$ (e.g., provided by an estimation procedure from a training set). 
In functional discriminant analysis, a new curve $\bx_i$ is assigned to the class $\hat{y}_i$ using the  maximum a posteriori (MAP) rule, that is: 
\begin{equation}
\hat{y}_i=\arg \max_{1\leq g\leq G} \frac{w_g p(\bx_i|y_i=g,\bt;\bsPsi_g)}{\sum_{g'=1}^{G}w_{g'}p(\bx_i|y_i=g',\bt;\bsPsi_{g'})} ,
\label{eq: MAP rule for FDA classification}
\end{equation}where $w_g = p(y_i=g)$ is the prior probability of class $g$, which can be computed as the proportion of the class $g$ in the training set, and $p(\bx_i|y_i=g,\bt;\bsPsi_g)$ its  conditional density. 

Different ways are possible to model this conditional density. 
By analogy to linear or quadratic discriminant analysis for vectorial data, the  class conditional density for each class of curves can be defined as a density of a single model, e.g., a polynomial regression model,  spline, including B-spline \citep{garetjamesANDtrevorhastieJRSS2001}, or a generative piecewise regression model with a hidden logistic process (RHLP)  \citep{chamroukhi_et_al_neurocomputing2010} when the curves further present regime changes over time. These approaches lead to Functional Linear (or quadratic) Discriminant Analysis which we will abbreviate as (FLDA).

In the next section, we briefly recall the FLDA based on polynomial or spline regression.

\subsection{Functional Linear Discriminant Analysis}
\label{ssec: FLDA state of the art}
Functional Linear  Discriminant Analysis (FLDA), firstly proposed in \cite{garetjamesANDtrevorhastieJRSS2001} for irregularly sampled curves, arises when we model each  class conditional density of curves with a single model. More specifically, the conditional density $p(\bx_i|y=g,\bt;\bsPsi_g)$ in Equation (\ref{eq: MAP rule for FDA classification}) can for example be the one of a polynomial, spline or B-spline regression model with parameters  $\bsPsi_g$, that is:
\begin{eqnarray}
p(\bx_i|y_i=g,\bt;\bsPsi_g) = \N (\bx_{i};\bT \bsbeta_g,\sigma_g^2\Identity_m), 
\end{eqnarray}where $\bsbeta_{g}$ is the coefficient  vector of the polynomial or spline regression model representing class $g$ and $\sigma_{g}^2$ the associated noise variance, the matrix $\bT$ is the matrix of design which depends on the adopted model (e.g., for polynomial regression, $\bT$ is the $m\times(p+1)$ Vandermonde matrix with rows $(1,t_j,t_j^2,\ldots,t_j^p)$ for $j=1,\ldots,m.$, $p$ being the polynomial degree), and $\N (.;\bsmu,\bsSigma)$ represents the multivariate Gaussian density with mean $\bsmu$ and covariance matrix $\bsSigma$.
In this case, estimating the model for each class consists therefore in estimating the regression model parameters $\bsPsi_g$, namely by maximum likelihood which is equivalent to performing least squares estimation  in this Gaussian case. 
A similar FLDA approach that fits a specific generative piecewise regression model governed by a hidden logistic process to homogeneous classes of curves presenting regime changes has been presented in \cite{chamroukhi_et_al_neurocomputing2010, chamroukhi_PhD_2010}.

However, all these approaches, as they involve a single component density model for each class, are only suitable for homogeneous classes of curves. For complex-shaped classes, when one or more classes are dispersed, the hypothesis of a single component density model description for the whole class of curves becomes restrictive. This problem can be handled, by analogy to mixture discriminant analysis for vectorial data \citep{hastieANDtibshiraniMDA}, by adopting a mixture model formulation \citep{mclachlanFiniteMixtureModels, titteringtonBookMixtures} in the functional space for each class  of curves. The functional mixture can for example be a polynomial regression mixture  or a spline regression mixture  \citep{GaffneyThesis, chamroukhi_PhD_2010, Gui_FMDA}. This leads to Functional Mixture Discriminant Analysis (FMDA) \citep{chamroukhi_esann2012, chamroukhi_PhD_2010, Gui_FMDA}.
  
  The next section describes the previous work on FMDA which uses polynomial regression and spline regression mixtures.

\subsection{Functional Mixture Discriminant Analysis with  polynomial regression and spline regression mixtures}
\label{FMDA from the state of the art}
 
A first idea on Functional Mixture Discriminant Analysis (FMDA), motivated by the complexity of the time course gene expression functional data for which modeling each class with a single function using FLDA is not adapted, was proposed by \cite{Gui_FMDA} and is based on B-spline regression mixtures.  
In the approach of \cite{Gui_FMDA}, each class $g$ of functions is modeled as a mixture of $K_g$ sub-classes, each sub-class  $k$ ($k=1,\ldots,K_g$) is a noisy  B-spline function (can also be a polynomial or  a spline function) with parameters $\bsPsi_{gk}$. 
The model is therefore defined by the following conditional mixture density: 
\begin{eqnarray}
p(\bx_i|y_i = g, \bt; \bsPsi_g) &=& \sum_{k=1}^{K_g} \alpha_{gk} \ p(\bx_i|y_i=g,z_{gi}=k, \bt;\bsPsi_{gk}) \nonumber \\
&=& \sum_{k=1}^{K_g} \alpha_{gk} \N (\bx_{i};\bT \bsbeta_{gk},\sigma_{gk}^2\Identity_m), 
\label{eq: class mixture density for classic FMDA}
\end{eqnarray}where the $\alpha_{gk}$'s are the non-negative mixing proportions that sum to 1   such that  $\alpha_{gk} = p(z_i = k|y_i=g)$ ($\alpha_{gk}$ represents the prior probability of the sub-class $k$ of class $g$),
 $z_i$ is a hidden discrete variable in $\{1,\ldots,K_g\}$ representing the labels of the sub-classes for each class, and $\Identity_m$ is the $m$ dimensional identity matrix.   The parameters of this functional mixture density  for each class $g$ (Equation (\ref{eq: class mixture density for classic FMDA})), denoted by 
$$\bsPsi_g = (\alpha_{g1},\ldots,\alpha_{gK_g}, \bsPsi_{g1},\ldots,\bsPsi_{gK_g})$$
can be estimated by maximizing the observed-data log-likelihood by using the expectation-maximization (EM) algorithm \citep{dlr, mclachlanEM} as in \cite{Gui_FMDA}.

However, using polynomial or spline regression for class representation, as studied in \cite{chamroukhi_PhD_2010, chamroukhi_et_al_neurocomputing2010} is more adapted for curves presenting smooth regime changes and for the splines the knots have to be fixed in advance.  
When the regime changes are abrupt, capturing the regime transition points needs to relax the regularity constraints on splines, since a spline is a smooth function \citep{deboor1978},
 which leads to piecewise regression \citep{McGee}  for which the knots can be optimized using a dynamic programming procedure \citep{bellman, stone}. On the other hand, the regression model with a hidden logistic process (RHLP) presented in \cite{chamroukhi_et_al_neurocomputing2010} and used to model each homogeneous set of curves with regime changes, is flexible and explicitly integrates the smooth and/or abrupt regime changes via a logistic process. As pointed in \cite{chamroukhi_et_al_neurocomputing2010}, this approach however has  limitations in the case of complex-shaped classes of curves since each class is only approximated by a single RHLP model.

In this paper, we  extend the discrimination approach proposed in \cite{chamroukhi_et_al_neurocomputing2010} which is based on functional linear discriminant analysis (FLDA) using  a single density model (RHLP) for each class, to a functional mixture discriminant analysis framework (FMDA), where each class conditional density model is assumed to be a specific mixture density. This density is a mixture of regression models with hidden logistic processes, which we will abbreviate as MixRHLP.  
Thus, by using this Functional Mixture Discriminant Analysis approach, We may therefore overcome the limitation of FLDA (and FQDA) for modeling complex-shaped classes of curves, via the mixture formulation. Furthermore,  thanks to the flexibility to the RHLP model that approximates each sub-class, as studied in \cite{chamroukhi_et_al_NN2009, chamroukhi_et_al_neurocomputing2010}, we will also be able to automatically and flexibly approximate the underlying hidden regimes for each sub-class.  

The proposed Functional Mixture Discriminant Analysis with hidden process regression and the unsupervised learning procedure  for each class through the EM algorithm, are presented in the next section. 

\section{Proposed Functional Mixture Discriminant Analysis with hidden process regression mixture}
\label{FMDA from the state of the art}
Let us assume as previously that each class $g$ 
has a complex shape so that it is composed of $K_g$ homogeneous sub-classes. Furthermore, now let us suppose that each sub-class $k$ 
of class $g$  is itself governed by $L_{gk}$ unknown regimes.

\subsection{Modeling the classes of curves with a mixture of regression models with hidden logistic processes} 
 \label{ssec: mixrhlp}
In the proposed Functional Mixture Discriminant Analysis (FMDA) approach, we model each class of curves by a specific mixture of Regression models with Hidden Logistic Processes (MixRHLP) as in \cite{chamroukhi_PhD_2010,chamroukhi_adac_2011}.   The approach will thus abbreviated as FMDA-MixRHLP. 
According to the MixRHLP model, each class of curves $g$ is assumed to be composed of $K_g$ homogeneous sub-groups with prior probabilities $\alpha_{g1},\ldots,\alpha_{gK_g}$. Each of the $K_g$ sub-groups is governed by $L_{gk}$ hidden polynomial regimes. 
Thus, for the $i$th curve $\bx_i$ issued from sub-class $k$ of class $g$, the observation $x_{ij}$ may switch from one regime to another at each time point $t_j$.

\bigskip
We let $z_{gi}$ denotes the variable representing the unobserved (hidden) sub-class (cluster) label of the $i$th curve $\bx_i$ for class $g$. We have therefore $z_{gi}= k \in \{1,\ldots,K_{g}\}$ for sub-class $k$. We will thus denote by $\bz_g=(z_{g1},\ldots,z_{gn})$ the hidden cluster labels for class $g$.
\\
Furthermore, we let $r_{gkj}$ denotes the unobserved regime label for sub-class $k$ of class $g$ at time $t_j$. Thus, we have $r_{gkj} = \ell \in \{1,\ldots,L_{gk}\}$ for regime $\ell$.
The variable $r_{gkj}$ allows for  switching from one regime to another among $L_{gk}$ regimes over time. We let therefore $\bh_{gk} = (h_{gk1},\ldots,h_{gkm})$ denotes the labels vector governing each sub-class $k$ of class $g$.  
\\
In the following, we will encode each of the random variables $z$ and $r$ in a binary manner as follows. 
The variable $z$ will be indexed by the three indexes
\begin{itemize}
\item $g$ : the group (class)
\item $k$ : the sub-class (cluster)
\item $i$ : the observation, which is the $i$th curve in this case,
\end{itemize}
so that we have $z_{gki}$ equals 1 if and only if the sub-class label of the curve $\bx_i$ belonging to class $g$ is $k$, that is $z_{gi} = k$.
Similarly, we will use a binary coding for the variable $r$ which will be indexed by the four indexes
\begin{itemize}
\item $g$ : the group (class)
\item $k$ : the sub-class (cluster)
\item $\ell$ : the regime
\item $j$ : the observation at time $t_j$,
\end{itemize}
so that we have $r_{gk\ell j}$ equals 1 if and only if, for sub-class $k$ of class $g$, the regime at time $t_j$ is $\ell$, that is $r_{gkj} = \ell$.

\bigskip
The distribution of each configuration of the discrete variable $r_{gkj}$ is assumed to be logistic, thus $\bh_{gk}$ governing each sub-class is therefore assumed to be a logistic process. This choice is due to the flexibility of the logistic function in both determining the regime transition points and accurately modeling abrupt and/or smooth regimes changes. Indeed, as it has been well detailed in \cite{chamroukhi_et_al_NN2009, chamroukhi_et_al_neurocomputing2010}, 
the logistic function (\ref{eq: logistic prob for regime g k r}) controls through its parameters the regime transition points and the quality of regime (smooth or abrupt) via the parameters $\{w_{gk\ell0},  w_{gk\ell1}\}$. The probability of each regime $\ell$ is given by: 
\begin{eqnarray} 
 \pi_{gk\ell}(t_j;\bw_{gk}) &=& p(r_{gkj}=\ell|t_j;\bw_{gk}) \nonumber \\
 &=& \frac{\exp{(w_{gk\ell 0} + w_{gk\ell 1}t_j)}}{\sum_{u=1}^{L_{gk}}\exp{(w_{gk \ell u 0} + w_{g k \ell u 1} t_j)}},
\label{eq: logistic prob for regime g k r}
\end{eqnarray}where $\bw_{gk} = (\bsw_{gk1},\ldots,\bsw_{gkL_{gk}})$ is its parameter vector, $\bsw_{gk\ell}=(w_{gk\ell0},w_{gk\ell1})$ being the $2$-dimensional coefficient vector for the $\ell$th logistic component. %
Furthermore, the regime are assumed to be noisy polynomial functions and the resulting model for each sub-class is the regression model with hidden logistic process (RHLP) \citep{chamroukhi_et_al_NN2009, chamroukhi_et_al_neurocomputing2010}. The RHLP model indeed assumes that the curves of each sub-class (or cluster) $k$ of class $g$ are generated by $K_g$ polynomial regression models governed by a hidden logistic process $\bh_{gk}$. 
Thus, as stated in \cite{chamroukhi_et_al_NN2009, chamroukhi_et_al_neurocomputing2010}, the distribution of a curve $\bx_i$ belonging to sub-class $k$ of class $g$ is defined by:
\begin{equation} 
p(\bx_i|y_i = g, z_{gi}=k,\bt;\bsPsi_{gk})  =  \prod_{j=1}^m \sum_{\ell=1}^{L_{gk}}\pi_{gk\ell}(t_j;\bw_{gk})\mathcal{N}\big(x_{ij};\bsbeta_{gk\ell}^T \bt_{j},\sigma_{gk\ell}^{2} \big) 
 \label{eq: RHLP}
\end{equation}
where $\bsPsi_{gk} = (\bw_{gk},\bsbeta_{gk1},\ldots,\bsbeta_{gkL_{gk}},\sigma^2_{gk1},\ldots,\sigma^2_{gkL_{gk}})$ for $(g=1,\ldots,G; k=1,\ldots,K_g)$
is its parameter vector.  
Hence, the resulting conditional distribution of a curve $\bx_i$ issued from class $g$ is given by the following conditional mixture density: 
{\small \begin{eqnarray}
p(\bx_i|y_i = g, \bt;\bsPsi_g) & =& \sum_{k=1}^{K_g} p(z_i = k|y_i = g) p(\bx_i|y_i = g, z_i\! =\! k,\bt;\bsPsi_{gk}) \nonumber \\
&=& \sum_{k=1}^{K_g} \alpha_{gk} \prod_{j=1}^m \sum_{\ell=1}^{L_{gk}}\pi_{gk\ell}(t_j;\bw_{gk})\mathcal{N}\big(x_{ij};\bsbeta_{gk\ell}^T \bt_{j},\sigma_{gk\ell}^{2} \big) 
\label{eq: MixRHLP}
\end{eqnarray}}where $\bsPsi_g =(\alpha_{g1},\ldots,\alpha_{gK_g},\bsPsi_{g1},\ldots,\bsPsi_{gK_g})$ is the parameter vector for class $g$,  
 $\bsPsi_{gk}$ being the parameters of each of its RHLP component density, that is $\prod_{j=1}^m \sum_{\ell=1}^{L_{gk}}\pi_{gk\ell}(t_j;\bw_{gk})\mathcal{N}\big(x_{ij};\bsbeta_{gk\ell}^T \bt_{j},\sigma_{gk\ell}^{2} \big)$ as given by Equation (\ref{eq: RHLP}). Notice that the key difference between the proposed FMDA with hidden process regression mixture  and the FMDA proposed in \cite{Gui_FMDA} is that the proposed approach uses a generative hidden process regression model (RHLP) for each sub-class rather than a spline; the RHLP is itself based on a dynamic mixture formulation as it can be seen in Equation (\ref{eq: RHLP}).  Thus, the proposed approach is more adapted for capturing the regime changes within curves during time.  

Now, once we have defined the model for each class of curves $g$, we have to estimate its parameters $\bsPsi_g$. The next section presents the unsupervised learning of the model parameters $\bsPsi_g$ for each class of curves by maximizing the observed-data log-likelihood through the EM algorithm.

\subsection{Maximum likelihood  estimation via the EM algorithm}
\label{sec: parameter estimation by EM mixture functional rhlp} 

Given an independent training set of labeled curves $((\bx_1,y_1),\ldots,(\bx_n,y_n))$, the parameter vector $\bsPsi_g$ of the mixture density of class $g$ given by Equation (\ref{eq: MixRHLP}) is estimated by maximizing the following observed-data log-likelihood:
{  \begin{eqnarray} 
\cL(\bsPsi_g) & = & \log \prod_{i|y_i =g}\!\! p(\bx_i|y_i = g, \bt;\bsPsi_g)\nonumber\\
& = & \!\!\sum_{i|y_i=g} \log  \sum_{k=1}^{K_g} \alpha_{gk}  \prod_{j=1}^m  \sum_{\ell=1}^{L_{gk}} \pi_{gk\ell}(t_j;\bw_{gk})\mathcal{N}\big(x_{ij};\bsbeta_{gk\ell}^T \bt_{j},\sigma_{gk\ell}^{2} \big).\nonumber
\label{eq: loglik MixFRHLP for class g}
\end{eqnarray}}The maximization of this log-likelihood cannot be performed in a closed form. We maximize it 
iteratively by using a dedicated EM algorithm. 
The EM scheme requires the definition of the complete-data log-likelihood. The complete-data log-likelihood for the proposed MixRHLP model for each class, given the observed data which we denote by $\mathcal{D}_g = \big(\{\bx_i|y_i=g\},\bt\big)$, the hidden cluster labels $\bz_g$, and the hidden processes $\{\bh_{gk}\}$ governing the $K_g$ clusters, is given by:
\vspace*{-.08cm}
{ \begin{eqnarray} 
 \cL_c(\bsPsi_g) & = &\sum_{i|y_i = g} \sum_{k=1}^{K_g}  z_{gki}  \Big[ \log \alpha_{gk}  +   \sum_{j=1}^{m}\sum_{\ell=1}^{L_{gk}}  r_{gk\ell j} \log \pi_{gk\ell}(t_j;\bw_{gk}) \nonumber \\
&  + &  \sum_{j=1}^{m} \sum_{\ell=1}^{L_{gk}} r_{gk\ell j} \log \mathcal{N}\left(y_{ij};{\bsbeta}^{T}_{gk\ell}\bt_{j},\sigma^2_{gk\ell}\right)\Big]. 
\label{eq: complete log-lik for the MixRHLP}
\end{eqnarray}} 

The next paragraph shows how the observed-data log-likelihood $\cL(\bsPsi_g)$ is maximized by the EM algorithm.

\subsection{The dedicated EM algorithm for the unsupervised learning of the parameters of the MixRHLP model for each class}
\label{ssec. EM algorithm for mixture functional rhlp}
For each class $g$, the EM algorithm starts with an initial parameter $\bsPsi_g^{(0)}$ and alternates between the two following steps until convergence:

\subsubsection{E-step}
\label{par: E-step mixture of rhlp and EM}
This step computes the expected complete-data log-likelihood, given the observations $\mathcal{D}_g$, and the current parameter estimation  $\bsPsi_g^{(q)}$, $q$ being the current iteration number: 
{  \begin{eqnarray}  
\!\! \!\!\!\! Q(\bsPsi_g,\bsPsi_g^{(q)}) & =& \E\left[\cL_c(\bsPsi_g)\big|\mathcal{D}_g;\bsPsi_g^{(q)}\right] \cdot \nonumber
\label{eq: Q-function}
\end{eqnarray}}
As it can be seen in the expression of $\cL_c(\bsPsi_g)$, this step simply requires the calculation of conditional expectations of the variables $z_{gki}$ and $r_{gk\ell j}$. More specifically, the expected complete-data log-likelihood is given by:
{  \begin{eqnarray} 
\!\! \!\!\!\! Q(\bsPsi_g,\bsPsi_g^{(q)}) & =& \E\left[\cL_c(\bsPsi_g)\big|\mathcal{D}_g;\bsPsi_g^{(q)}\right]\nonumber \\  
\!\! & = &\!
\sum_{i|y_i=g}\!\sum_{k=1}^{K} \!\!  \gamma_{gki}^{(q)} \log \alpha_{gk} \! + \!\!\!\! \sum_{i|y_i=g} \! \sum_{k=1}^{K_g}\! \sum_{j=1}^{m}\! \sum_{\ell=1}^{L_{gk}}\!\! 
\gamma_{gki}^{(q)}\tau^{(q)}_{gk\ell i j} 
\log \pi_{gk\ell}(t_j;\bw_{gk}) \nonumber \\ 
& &  + \!\! \sum_{i|y_i=g} \sum_{k=1}^{K_g} \sum_{j=1}^{m}\sum_{\ell=1}^{L_{gk}} 
\gamma_{gki}^{(q)} \tau^{(q)}_{gk\ell i j} \log \mathcal{N} \left(x_{ij};{\bsbeta}^{T}_{gk\ell}\bt_{j},\sigma^2_{gk\ell} \right).\label{eq: Q-function for the MixRHLP for class g}
\end{eqnarray}}
As shown in the expression of $Q(\bsPsi_g,\bsPsi_g^{(q)})$, this step simply requires the calculation of the posterior sub-class probabilities, i.e., the probability that the observed curve $\bx_{i}$ originates from sub-class (cluster) $k$ for class $g$, which we index in a similar way as $z_{gki}$ and are given by: 
{  \begin{eqnarray}
\gamma_{gki}^{(q)} 
& = & p(z_{gi}=k|\bx_{i},y_i=g,\bt;\bsPsi_{gk}^{(q)}) \nonumber \\
&=& \frac{\alpha_{gk}^{(q)} p(\bx_i | y_i=g,z_{gi}=k,\bt;\bsPsi^{(q)}_{gk})}{ \sum_{k\prime  =1}^{K_g} \alpha_{gk' }^{(q)}p(\bx_i |y_i=g, z_{gi}=k' ,\bt;\bsPsi^{(q)}_{gl})} \nonumber \\ 	
&=& \frac{\alpha_{gk}^{(q)}\prod_{j=1}^m\sum_{\ell=1}^{L_{gk}}\pi_{gk\ell}(t_j;\bw_{gk}^{(q)})\mathcal{N}\big(x_{ij};\bsbeta^{T(q)}_{gk\ell}\bt_{j},\sigma^{2(q)}_{gk\ell}\big)}
{ \sum_{k' =1}^{K_g} \alpha_{gk'}^{(q)}\prod_{j=1}^m\sum_{\ell=1}^{R_{gk'}} \pi_{gk'r}(t_j;\bw_{gk'}^{(q)}) \mathcal{N}(x_{ij};\bsbeta^{(q)T}_{gk' \ell}\bt_{j},\sigma^{2(q)}_{gk' \ell})}    
\label{eq: curves post prob gamma_ijgk}
\end{eqnarray}}and the posterior regime probabilities for each sub-class, i.e., the probability that the observed data point $x_{ij}$ at time $t_j$ originates from the $\ell$th regime of sub-class $k$ for class $g$, which we index in a similar way as $r_{gk\ell j}$ and are given by:
\begin{eqnarray}
\tau^{(q)}_{gk\ell i j}  
&=& p(r_{gkj}=\ell|x_{ij},y_i=g, z_{gi}=k,  t_j;\bsPsi^{(q)}_{gk})\nonumber \\ 
&=&\frac{\pi_{gk\ell}(t_j;\bw_{gk}^{(q)})\mathcal{N}(x_{ij};\bsbeta^{T(q)}_{gk\ell}\bt_{j},\sigma^{2(q)}_{gk\ell})}
{\sum_{l=1}^{L_{gk}}\pi_{gkl}(t_j;\bw_{gk}^{(q)})\mathcal{N}(x_{ij};\bsbeta^{T(q)}_{gk l}\bt_{j},\sigma^{2(q)}_{gk l})}\cdot
\label{eq: post prob tau^r_ijk of the segment k of the cluster r for the MixRHLP}
\end{eqnarray}

\subsubsection{M-step}
\label{par: M-step mixture of rhlp and EM}
This step updates the value of the parameter $\bsPsi_g$ by maximizing the function $Q(\bsPsi_g,\bsPsi_g^{(q)})$ given by Equation (\ref{eq: Q-function for the MixRHLP for class g}) with respect to $\bsPsi_g$, that is: 
$$\bsPsi_g^{(q+1)} = \arg \max_{\bsPsi_g} Q(\bsPsi_g,\bsPsi_g^{(q)}).$$  
It can be shown that this maximization can be performed by separate  maximizations w.r.t the mixing proportions $(\alpha_{g1},\ldots,\alpha_{gK_g})$ subject to the constraint $\sum_{k=1}^{K_g} \alpha_{gk} = 1$, 
and w.r.t the regression parameters $\{\bsbeta_{gk\ell},\sigma^2_{gk\ell}\}$ and the hidden logistic processes' parameters $ \{\bw_{gk}\}$. 

The mixing proportions updates are given, as in the case of standard mixtures, by 
 \begin{eqnarray}
\alpha_{gk}^{(q+1)} &=& \frac{1}{n_g}\sum_{i|y_i=g} \gamma_{gki}^{(q)} \, ,  \quad (k=1,\ldots,K_g), 
\label{eq: EM estimate of the cluster prior prob alpha_r for the MixRHLP}
\end{eqnarray} $n_g$ being the cardinal number of class $g$. 
The maximization w.r.t the regression parameters consists in performing separate analytic solutions of weighted least-squares problems where the weights are the product of the posterior probability $\gamma^{(q)}_{gki}$  
 of sub-class $k$ and the posterior probability $\tau^{(q)}_{gk\ell i j}$  
 of regime $\ell$ of sub-class $k$. Thus, the regression coefficients updates are given by:
{ \begin{eqnarray}
 \bsbeta^{(q+1)}_{gk\ell} & = &\Big[\sum_{i|y_i=g}  \sum_{j=1}^{m} \gamma_{gki}^{(q)} \tau^{(q)}_{gk\ell i j}\bt_j \bt_j^T\Big]^{-1} \sum_{i|y_i = g} \sum_{j=1}^{m} \gamma_{gki}^{(q)} \tau^{(q)}_{gk\ell i j}x_{ij} \bt_j   
\label{eq: EM estimate of reg coeff beta_rk of polynom k of the sub-class r for the MixRHLP}
\end{eqnarray}
and the updates for the variances are given by:
\begin{eqnarray}
 \sigma_{gk\ell}^{2(q+1)} & \!\! = \!\! & \frac{\sum_{i|y_i=g}\sum_{j=1}^{m} \gamma_{gki}^{(q)} \tau^{(q)}_{gk\ell i j} (x_{ij}-{\bsbeta}^{T(q+1)}_{gk\ell} \bt_j)^2}{\sum_{i|y_i=g}\sum_{j=1}^m  \gamma_{gki}^{(q)} \tau^{(q)}_{gk\ell i j} } \cdot  
\label{eq: EM estimate of variance sigma^2_rk for polynom k of the sub-class r for the MixRHLP}
\end{eqnarray}}Finally, the maximization w.r.t the logistic processes parameters $\{\bw_{gk}\}$ consists in solving multinomial logistic regression problems weighted by  $\gamma_{gki}^{(q)}\tau^{(q)}_{gk\ell i j}$  which we solve with a multi-class IRLS algorithm.
The IRLS algorithm (e.g., see  \cite{irls, chamroukhi_PhD_2010}) is an iterative algorithm which consists of starting with an initial paramter vector $\bw_{gk}^{(0)}$, and updating the estimation until convergence. A single update of the IRLS algorithm at iteration $l$ is given by:
{ \begin{equation}
 {\bw}_{gk}^{(l+1)} = {\bw}_{gk}^{(l)}-\Big[\frac{\partial^2 Q^{(q)}_{\bw_{gk}})}{\partial \bw_{gk} \partial {\bw_{gk}}^T}\Big]^{-1}_{\bw_{gk}=\bw_{gk}^{(l)}} \frac{\partial Q^{(q)}_{\bw_{gk}}}{\partial \bw_{gk}}\Big|_{\bw_{gk}=\bw_{gk}^{(l)}}.
\label{eq: IRLS MixRHLP}
\end{equation}}where $Q^{(q)}_{\bw_{gk}}$ denotes the term in the $Q$-function (\ref{eq: Q-function for the MixRHLP for class g}) that depend on $\bw_{gk}$, that is $\sum_{i|y_i=g} \sum_{k=1}^{K_g} \sum_{j=1}^{m} \sum_{\ell=1}^{L_{gk}} \gamma_{gki}^{(q)} \tau^{(q)}_{gk\ell i j} \log \pi_{gk\ell}(t_j;\bw_{gk})$. 
The parameter update $ {\bw}_{gk}^{(q+1)} $ is then taken at convergence of the IRLS algorithm (\ref{eq: IRLS MixRHLP}).

The pseudo code \ref{algo: proposed EM algorithm for mixture of functional rhlps} summarizes the EM algorithm for the proposed MixRHLP model for each class.
\begin{algorithm}
\caption{\label{algo: proposed EM algorithm for mixture of functional rhlps} Pseudo code of the proposed algorithm for the MixRHLP model for a set of curves.} 
{\bf Inputs:} Labeled training set of curves $((\bx_1,y_1),\ldots,(\bx_n,y_n))$ sampled at the time points $\bt=(t_1,\ldots,t_m)$, the number of sub-classes $K_g$, the number of regimes $L_{gk}$ and the polynomial degree $p$.
\begin{algorithmic}[1] 
\STATE \textbf{Initialize:} $\bsPsi_g^{(0)}= (\alpha^{(0)}_{g1},\ldots,\alpha^{(0)}_{gK_g},\bsPsi_{g1}^{(0)},\ldots,\bsPsi_{gK_g}^{(0)})$ 
\STATE fix a threshold $\epsilon>0$ (e.g., $\epsilon=10^{-6}$), 
\STATE set $q \leftarrow 0$ (EM iteration) 
\WHILE {increment in log-likelihood $> \epsilon$}
\STATE \begin{verbatim}
// E-Step
\end{verbatim} 
	   \FOR{$k=1,\ldots,K_g$}		
	   	\STATE compute  $\gamma_{gki}^{(q)}$  for $i=1,\ldots,n$ using Equation (\ref{eq: curves post prob gamma_ijgk})

			\FOR{$r=1,\ldots,L_{gk}$}
				\STATE compute  $\tau_{gk\ell i j}^{(q)}$  for $i=1,\ldots,n$ ; $j=1,\ldots,m$ using Equation (\ref{eq: post prob tau^r_ijk of the segment k of the cluster r for the MixRHLP})
			\ENDFOR
	 \ENDFOR
\STATE \begin{verbatim}
// M-Step
\end{verbatim} 
	\FOR{$k=1,\ldots,K_g$}	
		   	\STATE compute the update $\alpha_{gk}^{(q+1)}$ using Equation (\ref{eq: EM estimate of the cluster prior prob alpha_r for the MixRHLP})	
		\FOR{$r=1,\ldots,L_{gk}$}
			\STATE compute the update $\bsbeta_{gk\ell}^{(q+1)}$ using Equation (\ref{eq: EM estimate of reg coeff beta_rk of polynom k of the sub-class r for the MixRHLP})
			\STATE compute the update $\sigma_{gk\ell}^{2(q+1)}$ using Equation (\ref{eq: EM estimate of variance sigma^2_rk for polynom k of the sub-class r for the MixRHLP})
		\ENDFOR 
		\STATE  \begin{verbatim}
		//IRLS updating loop (Eq. (14))
		\end{verbatim}
		\STATE \textbf{Initialize:} set $\bw_{gk}^{(l)} = \bw_{gk}^{(q)}$
		\STATE set a threshold $\zeta>0$
		\STATE  $l \leftarrow 0$ (IRLS iteration)
		\WHILE{increment in $Q_{\bw_{gk}}>\zeta$}
			\STATE compute $ \bw_{gk}^{(l+1)}$ using  Equation (\ref{eq: IRLS MixRHLP})
			\STATE $l \leftarrow l+1$
		\ENDWHILE
	\STATE$\bw_{gk}^{(q+1)} \leftarrow \bw_{gk}^{(l)}$
	\STATE $q \leftarrow q+1$
	  \ENDFOR
	\ENDWHILE
\STATE $\hat{\bsPsi}= (\alpha^{(q)}_{g1},\ldots,\alpha^{(q)}_{gK_g}, \bsPsi^{(q)}_{g1},\ldots \bsPsi^{(q)}_{gK_g})$\end{algorithmic}
{\bf Output:} $\hat{\bsPsi}_g$ the maximum likelihood estimate of $\bsPsi_g$ 
\end{algorithm}

\subsection{Curve classification and approximation with the FMDA-MixRHLP approach} 
\label{ssec: FMDA-MixRHLP classification and approximation}
This section relates to the approximation of each class of curves by a single or several curve models in the cas of a dispersed class, and the class prediction for new observed curves based on the learned classes parameters. Once we have an estimate $\hat{\bsPsi}_{g}$ of the parameters of the functional mixture density MixRHLP (provided by the EM algorithm) for each class, a new curve $\bx_i$ is then assigned to the class maximizing the posterior probability (MAP principle) using Equation (\ref{eq: MAP rule for FDA classification}). This therefore leads us to the functional mixture discriminant analysis classification rule (FMDA-MixRHLP) which is particularly adapted to deal with the problem of classes composed of several sub-classes and to further handle the problem of regime changes within each sub-class.

Concerning the curves approximation, each sub-class $k$ of class $g$ is summarized by approximating it by a single ``mean" curve, which we denote by $\hat{\bx}_{gk}$. Each point $\hat{x}_{gkj}   \ (j=1\ldots,m)$ of this mean curve is defined by the conditional expectation $\hat{x}_{gkj}=\E[x_{ij}|y_i=g,z_{gi}=k,t_j;\hat{\bsPsi}_{gk}]$ given by:  
\begin{eqnarray} 
\hat{x}_{gkj} &=& \int_{\R}x_{ij} p(x_{ij}|y_i=g,z_{gi}=k,t_j;\hat{\bsPsi}_{gk})dx_{ij} \nonumber \\
&=&\int_{\R}x_{ij} \sum_{\ell=1}^{L_{gk}} \pi_{gk\ell}(t_j;\hat{\bw}_{gk}) \mathcal{N}\big(x_{ij};\hat{\bsbeta}^T_{gk\ell}\bt_{j},\hat{\sigma}^2_{gk\ell}\big) dx_{ij} \nonumber \\
&=& \sum_{\ell=1}^{L_{gk}} \pi_{gk\ell}(t_j;\hat{\bw}_{gk})\hat{\bsbeta}^T_{gk\ell} \bt_{j} 
\label{eq: RHLP mean curve}
\end{eqnarray}which is a sum of polynomials weighted by the logistic probabilities $\pi_{gk\ell}$ that model the regime variability over time.

\subsection{Model selection}
\label{ssec: model selection method}

The number of sub-classes (clusters) $K_g$ for each class $g$ $(g=1,\ldots,G)$ and the number regimes $L_{gk}$ for each sub-class can be computed by maximizing some information criteria e.g., the Bayesian Information Criterion (BIC) \citep{BIC}:
\begin{equation}
\mbox{BIC}(K,R,p)=\cL(\hat{\bsPsi_g})-\frac{\nu_{\bsPsi_g}}{2} \log(n),
\label{eq: BIC for MixFRHLP}
\end{equation}
where $\hat{\bsPsi_g}$  is the maximum likelihood estimate of the parameter vector $\bsPsi_g$ provided by the EM algorithm, $\nu_{\bsPsi_g} = K_g-1 + \sum_{k=1}^{K_g} \nu_{\bsPsi_{gk}}$ is the number of free parameters of the MixRHLP model, $K_g-1$ being the number of mixing proportions and $\nu_{\bsPsi_{gk}} = (p+4)L_{gk}-2$  represents the number of free parameters of each RHLP model associated with sub-class $k$, and $n$ is the number of curves belonging to the training set of the considered class. 
Note that in  \cite{Gui_FMDA} the number of sub-classes are fixed by the user. 

In practice, the model selection procedure consists in specifying a maximum values of $(K_{max},R_{max},p_{max})$ and then running the EM algorithm on each class of curves for $k=1,\ldots,K_{max}$, $R=1,\ldots,R_{max}$
 and $p=0,\ldots,K_{max}$ and the corresponding BIC value is stored. The triplet corresponding to the highest value of BIC is then selected.
These computations for selecting three values can be computationally more expensive compared the ones in classical model selection namely for standard mixture where only the number of cluster has to be selected. However, we notice that for small values of $(K, R, p)$, the computational cost is around few minutes and is not dramatically high, compared to approaches involving dynamic programming namely when using piecewise regression or when training approaches requiring MCMC sampling. 
Furthermore, it can be noticed that in some real situations, such as the one we present later, one or more values can be known (i.e., fixed by the experts namely the number of regimes and the structure of regimes for which a three-polynomial degree is well adapted) and the corresponding model selection procedure is quite fast.   

The next section, we evaluate the proposed approach and perform comparisons with alternative ones.

\section{Experimental study}  
\label{sec: Experiments} 

This section is dedicated to the evaluation of the proposed approach. We tested it on simulated data, the waveform benchmark curves of Breiman \citep{breiman} and real data from a railway diagnosis application \citep{chamroukhi_et_al_NN2009,chamroukhi_et_al_neurocomputing2010,chamroukhi_adac_2011}.

 We performed comparisons with alternative functional discriminant analysis approaches  using a polynomial regression (PR) or a spline regression (SR) model \citep{garetjamesANDtrevorhastieJRSS2001}, and the one that uses a single RHLP model per class as in \cite{chamroukhi_et_al_neurocomputing2010}. These alternatives will be abbreviated as FLDA-PR, FLDA-SR and FLDA-RHLP, respectively.  We also considered alternative Functional Mixture Discriminant Analysis approaches that use polynomial regression mixtures (PRM), and spline regression mixtures (SRM) as in \cite{Gui_FMDA} which will be abbreviated as FMDA-PRM and FMDA-SRM respectively.

We used two evaluation criteria. The first one is the misclassification error rate  computed by a $5$-fold cross-validation procedure and concerns the performance of the approaches in terms of curve classification. The second one is the square error between the observed curves  and the estimated mean curves, which is equivalent to the intra-class inertia, and regards the performance of the approaches with respect to the curves modeling and approximation. For FLDA approaches, as each class $g$ is approximated by a single mean curve $\hat{\bx}_{g}$, this error criterion is therefore given by $\sum_{g}\sum_{i|y_i=g} \parallel \bx_{i} - \hat{\bx}_{g}\parallel^2$. While, for FMDA approaches, each class $g$ is summarised by several ($K_g$) mean curves $\{\hat{\bx}_{gk} \}$, each of them summarises a sub-class $k$, and the intra-class inertia in this case is therefore given by $\sum_{g}\sum_{k=1}^{K_g} \sum_{i|y_i=g,z_{gi}=k} \parallel \bx_{i} - \hat{\bx}_{gk}\parallel^2$. Notice that each point of the estimated mean curve for each sub-class is given by a polynomial function or a spline function for the case of polynomial regression mixture classification (FMDA-PRM) or spline regression mixture classification (FMDA-SRM) respectively, or by Equation (\ref{eq: RHLP mean curve}) for the case of the proposed FMDA-MixRHLP approach.  

\subsection{Experiments on simulated curves} 
\label{ssec: experiments on simulated data}
\begin{figure}[!h]
 \centering
 \includegraphics[width=5.6cm, height=3.8cm]{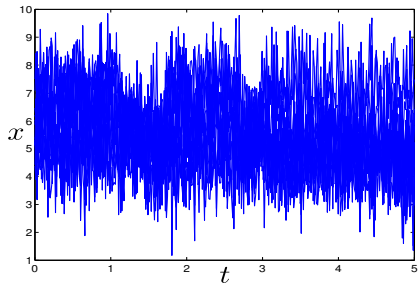} 
 \caption{\label{fig: simulated complex shaped class}
Simulated curves from a complex-shaped class composed of three sub-classes, each of them is composed of three constant regimes.}
\end{figure}

In this section, we consider simulated curves issued from two classes of piecewise noisy functions. The first class has a complex shape and is composed of three sub-classes (see Figure \ref{fig: simulated complex shaped class}), while the second one is a homogeneous class. Each sub-class is composed of 50 curves and each curve consists of three regimes and is composed of $200$ points.

\begin{figure}[!h]
 \centering 
 \includegraphics[width=5cm, height=3.3cm]{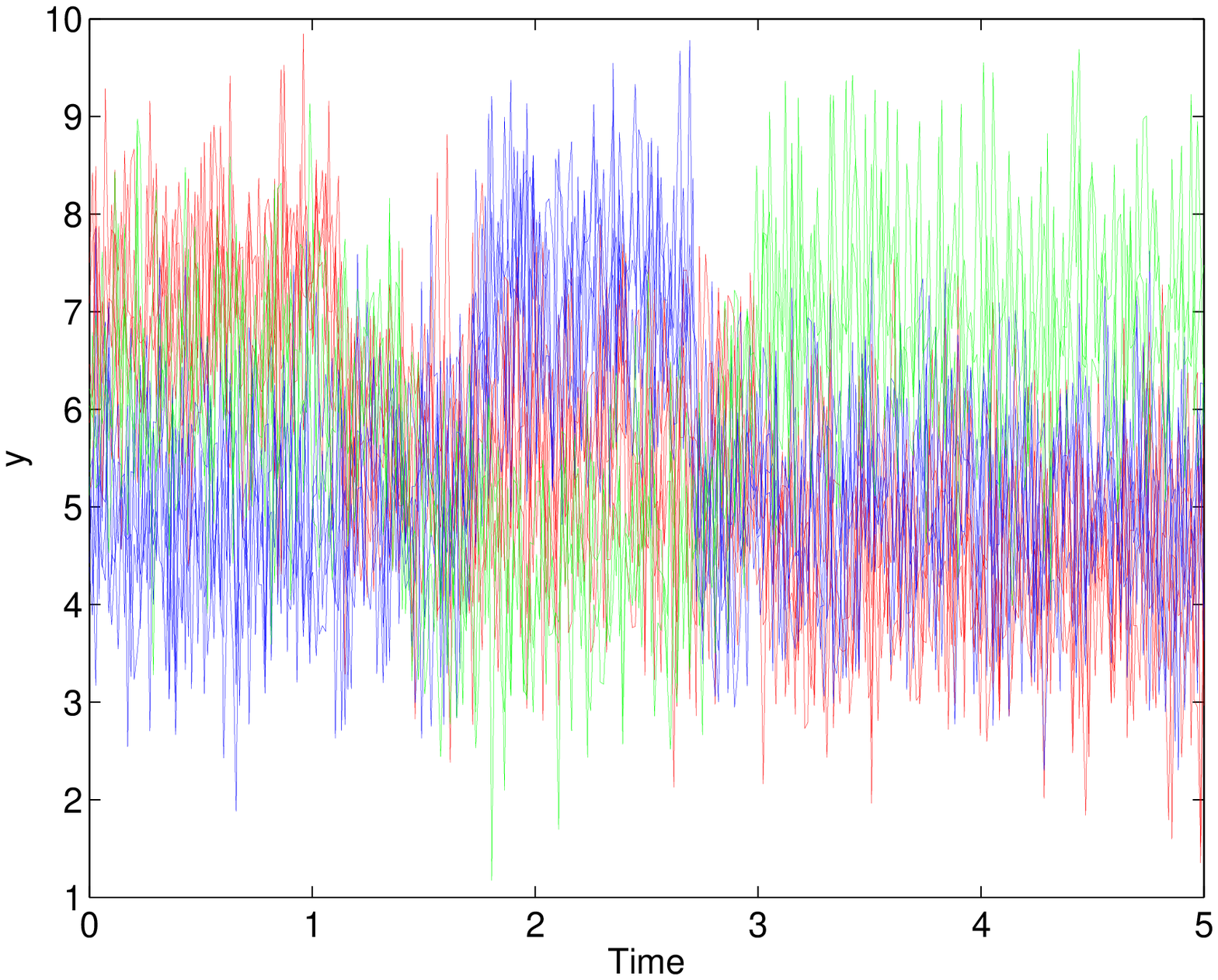} \\
 \includegraphics[width=4.48cm, height=4.3cm]{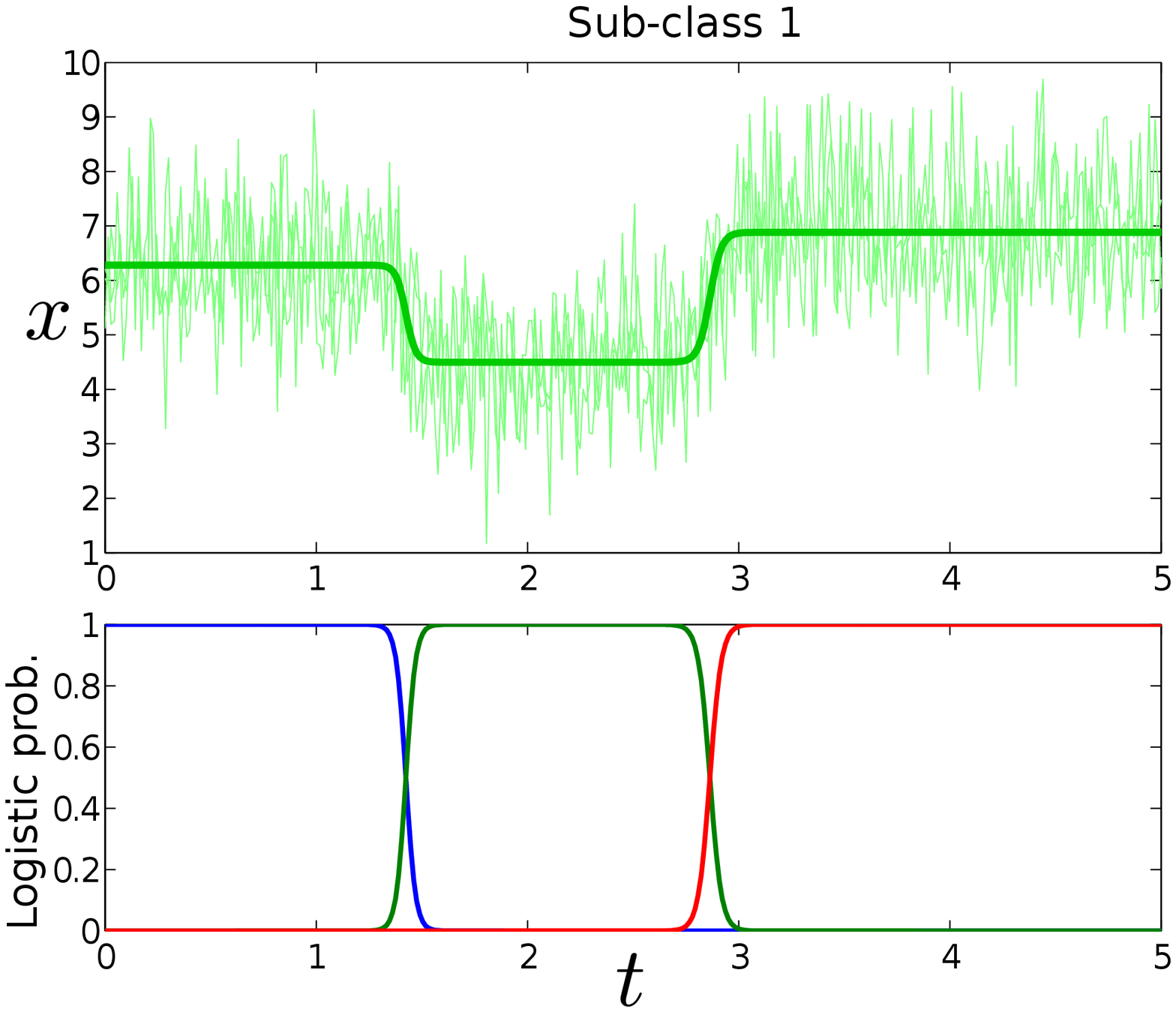} 
 \includegraphics[width=4.48cm, height=4.3cm]{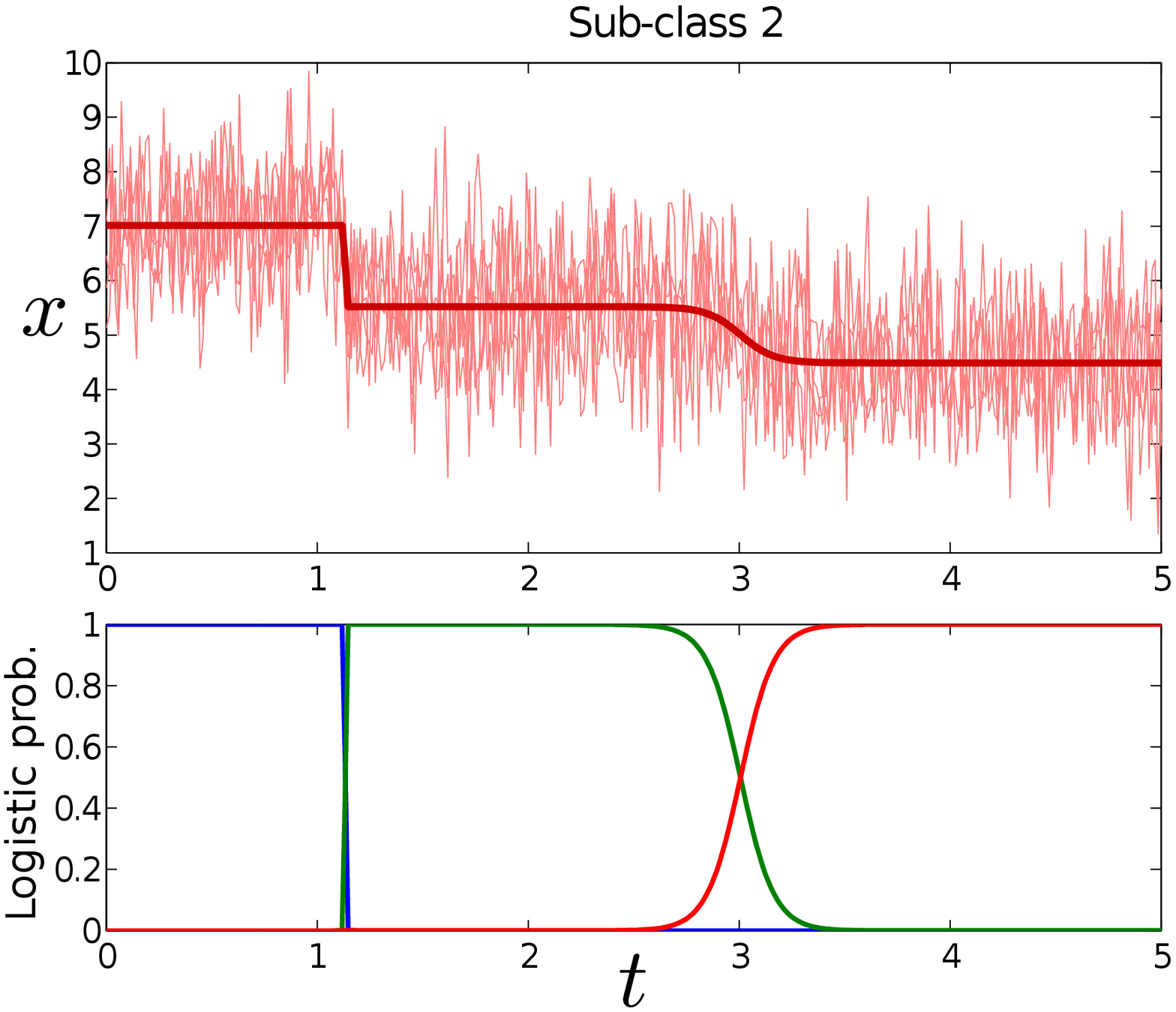}
 \includegraphics[width=4.48cm, height=4.3cm]{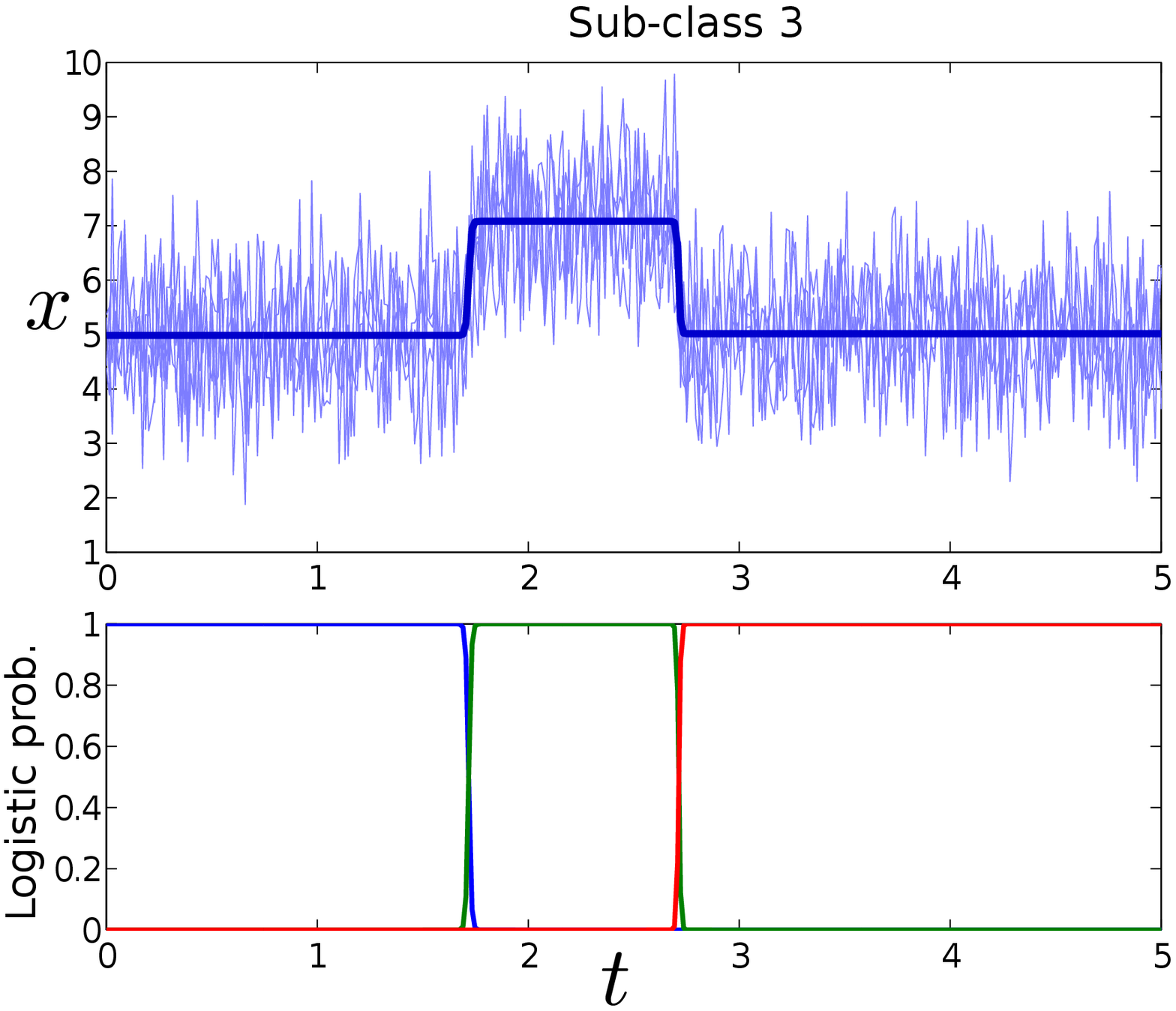}
\caption{\label{fig: results for the complex-shaped class}
The estimated sub-classes colored according to the partition given by the EM algorithm for the proposed approach (top); Then are presented separately each sub-class of curves with the estimated mean curve in bold line (top sub-plot) and the corresponding logistic probabilities that govern the hidden regimes (bottom sub-plot).} 
\end{figure}

Figure \ref{fig: results for the complex-shaped class} shows the obtained modeling results for the complex-shaped class shown in Figure \ref{fig: simulated complex shaped class}. 
It can be observed that the proposed approach accurately decomposes the class into homogeneous sub-classes of curves by automatically determining the sub-classes and the underlying hidden regimes for each sub-class.
Furthermore, the flexibility of the logistic process used to model the hidden regimes allows for accurately approximating both abrupt and/or smooth regime changes within each sub-class. This can be clearly seen on the logistic probabilities which vary over time according to both which regime is active or not and how is the transition from one regime to another over time (i.e., abrupt or smooth transition from one regime to another). We also notice that, approximating this class with a single mean curve, which is the case when using FLDA approaches (i.e., FMDA-PR or FMDA-SR), fails; the class is clearly heterogeneous. Using FMDA approaches based on polynomial or spline regression mixture (i.e., FMDA-PRM or FMDA-SRM) does not provide significant modeling improvements. This is due the fact that, as we can clearly see it on the data, the sub-classes present abrupt and smooth regime changes for which these two approaches are not well adapted. 
This is confirmed on the obtained intra-class inertia results given in Table \ref{table: results for simulated curves}. 
\begin{table}[!h]
\centering
\begin{tabular}{|l|c|c|} 
\hline
Approach &  Classification error rate (\%) & Intra-class inertia\\ 
\hline
\hline
FLDA-PR   & 21   &  $7.1364 \times 10^3$ \\
FLDA-SR  & 19.3 &  $6.9640 \times 10^3$ \\
FLDA-RHLP & 18.5 &  $6.4485\times 10^3$ \\
\hline
FMDA-PRM & 11 &    $6.1735 \times 10^3$\\
FMDA-SRM     & 9.5 & $5.3570 \times 10^3$ \\
{\bf FMDA-MixRHLP } & {\bf 5.3} & ${\bf 3.8095\times 10^3}$ \\
\hline
\end{tabular}
\caption{\label{table: results for simulated curves}
Obtained results for the simulated curves.}
\end{table} 
Table \ref{table: results for simulated curves} indeed shows that the smallest value of intra-class inertia is obtained for the proposed FMDA-MixRHLP approach. 
The proposed functional mixture discriminant analysis approach based on hidden logistic process regression (FMDA-MixRHLP) outperforms the alternative FMDA based on polynomial regression mixtures (FMDA-PRM) or spline regression mixtures (FMDA-SRM). This performance is attributed to the flexibility of the MixRHLP model thanks to the logistic process which is well adapted for modeling the regime changes. 
We cal also observe on Table \ref{table: results for simulated curves} that, as expected, the FMDA approaches outperforms the FLDA approaches. This is attributed to the fact that, in this case of heterogeneous class, FLDA  provides a rough class approximation  

Table \ref{table: results for simulated curves} also shows the misclassification error rates obtained with the proposed FMDA-MixRHLP approach and the alternative approaches.   
It can be seen that, also in terms of curve classification the FMDA approaches provide better results compared to FLDA approaches. This is due to the fact that using a single model for complex-shaped classes (i.e., when using FLDA approaches) is not adapted as it does not take into account the class dispersion when modeling the class conditional density. On the other hand, the proposed FMDA-MixRHLP approach provides a better modeling and therefore a more accurate class prediction.

\bigskip

We also performed experiments to select the best model for this dataset. 
The true values for the dispersed class are $(K=3, R=3, p=0)$ and for the other class which is homogeneous are $(K=1, R=3, p=0)$. 
For the procedure of model selection as described in section \ref{ssec: model selection method}, the values of $(K_{max},R_{max},p_{max})$ were set to $(K_{max}=4, R_{max}=4,p_{max}=4)$ and the proposed EM algorithm were applied to select the best model according to the highest BIC values. 
The process of model selection was repeated for 100 random samples.

 The percentage of choosing the best model for the first class composed of three sub-classes  is 91 \% while only 9 \% were obtained for the model $(K=3, R=3, p=1)$. This is attributed to the fact that the constant regimes may be approximated as well by a polynomial of order 1 (linear function). The percentage of choosing the best model for the second homogeneous class  is equal to 94 \%, the model corresponding to $(K=1, R=3, p=1)$ were selected in only 6 \% of cases.

\bigskip 

In the next section, the proposed approach  is applied on the waveform curves of Breiman \citep{breiman}.
\subsection{Waveform curves of Breiman}
\label{ssec: experiments using waveform data}

\begin{figure}[!h]
 \centering
\includegraphics[width=4.27cm,height=3.3cm]{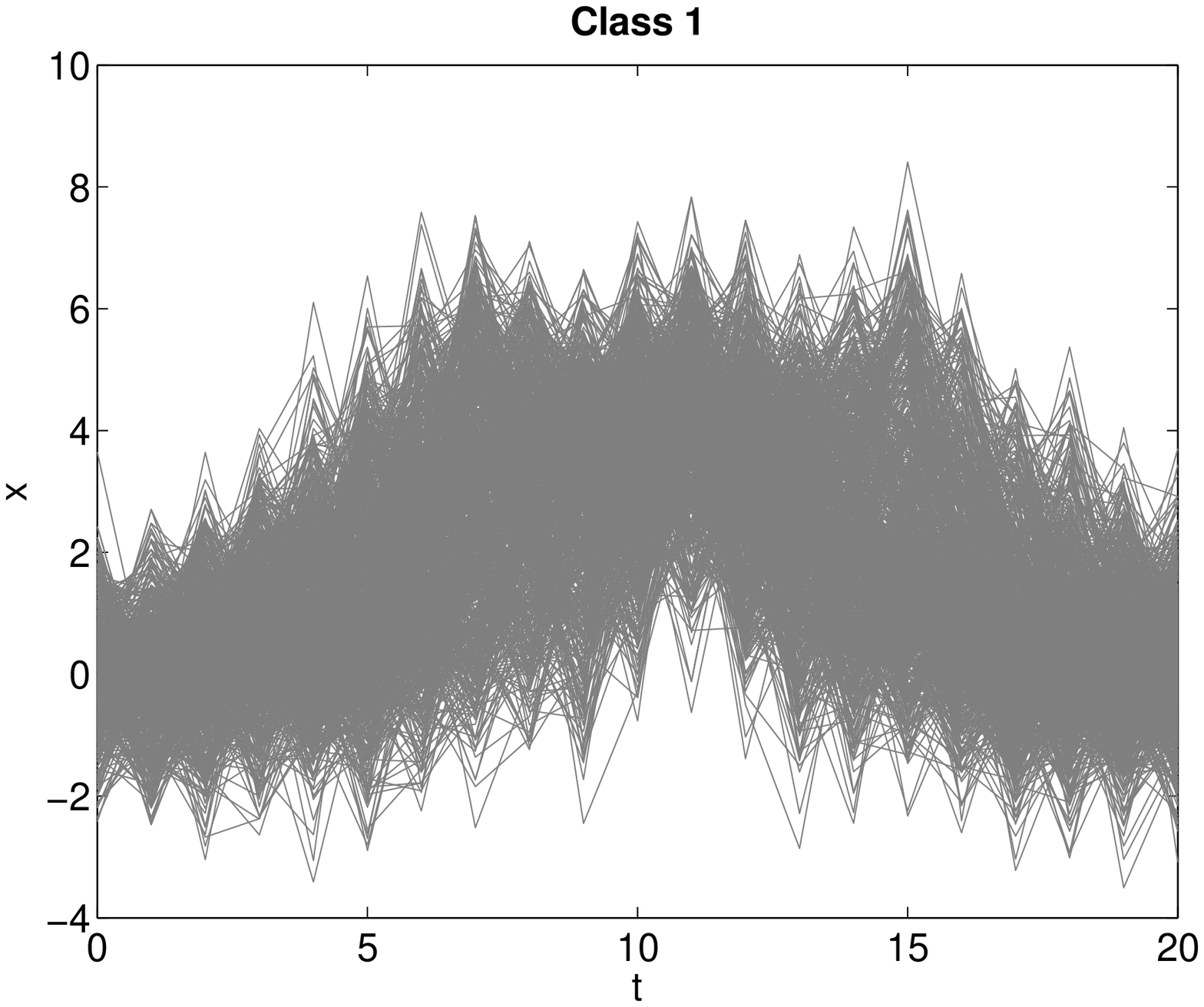}
 \includegraphics[width=4.27cm,height=3.3cm]{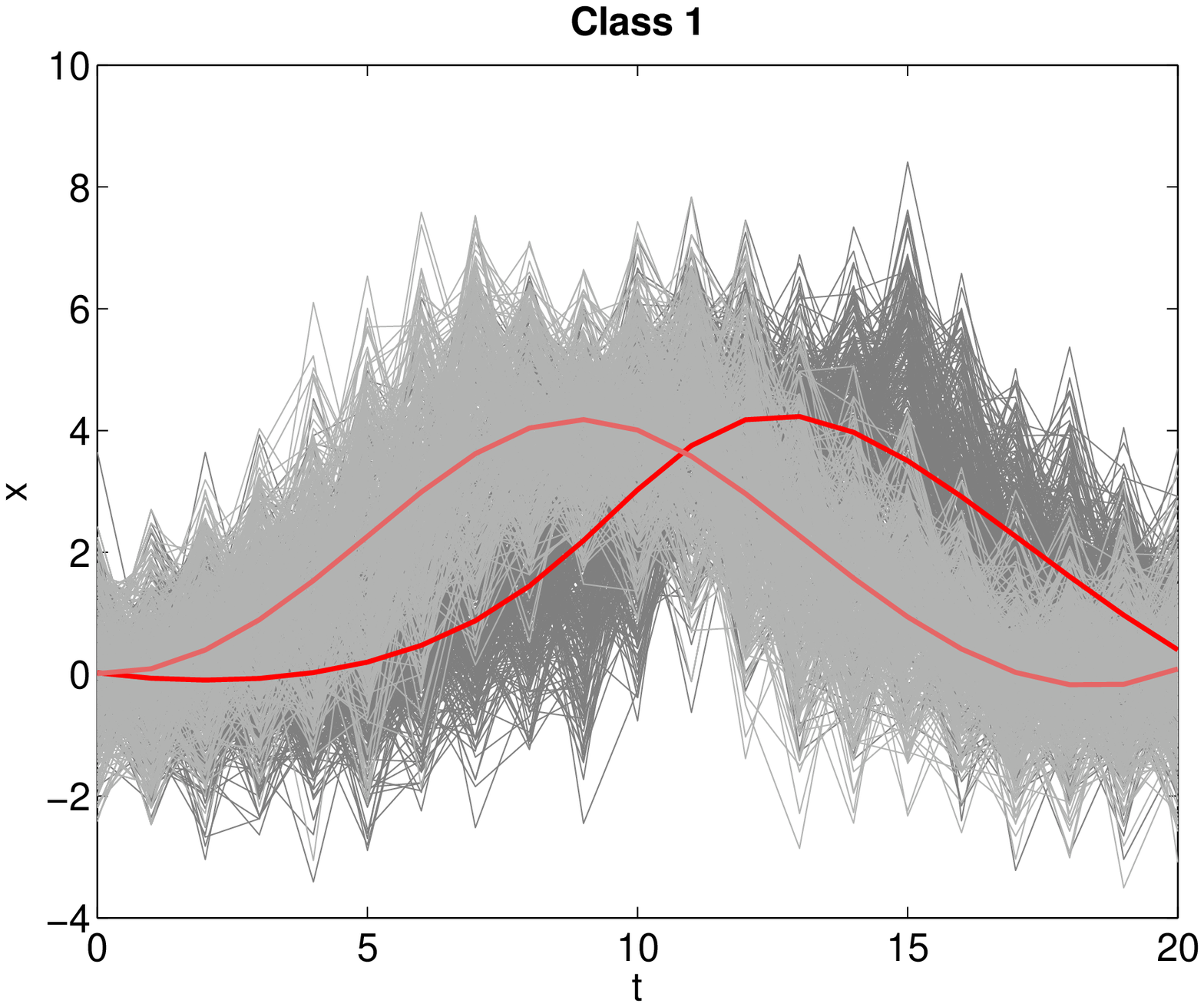}
\includegraphics[width=4.27cm,height=3.3cm]{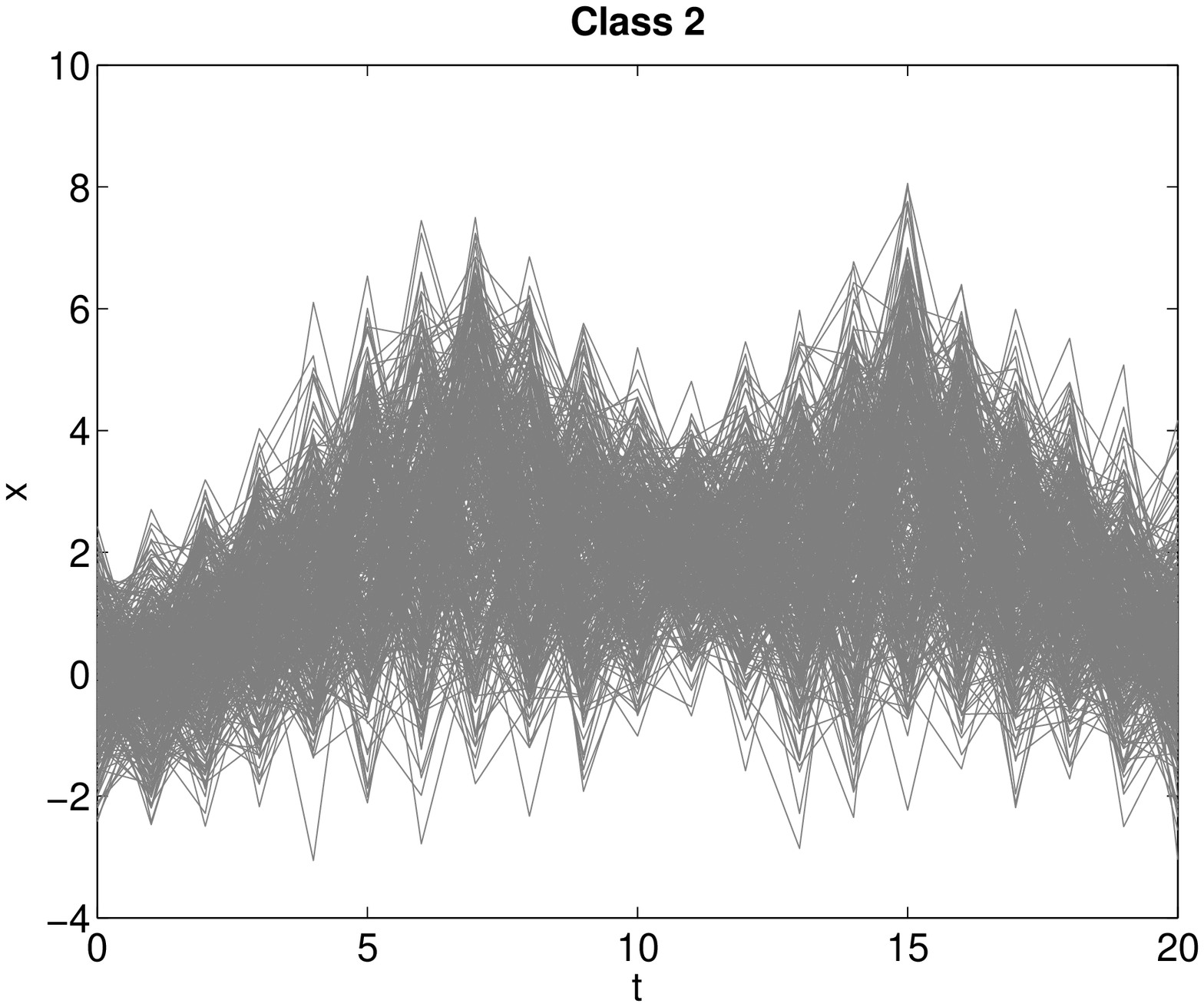}
\includegraphics[width=4.27cm,height=3.3cm]{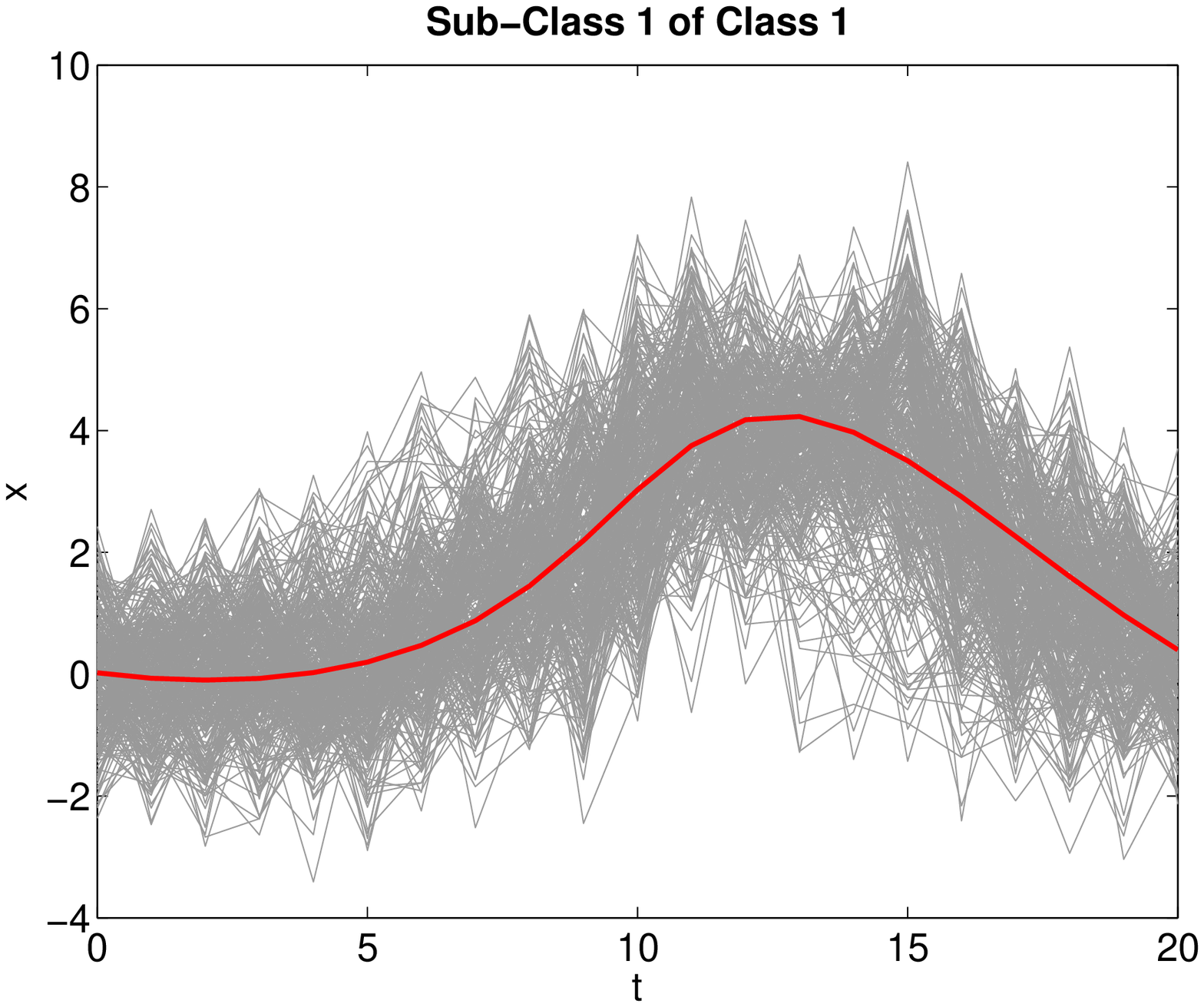}
\includegraphics[width=4.27cm,height=3.3cm]{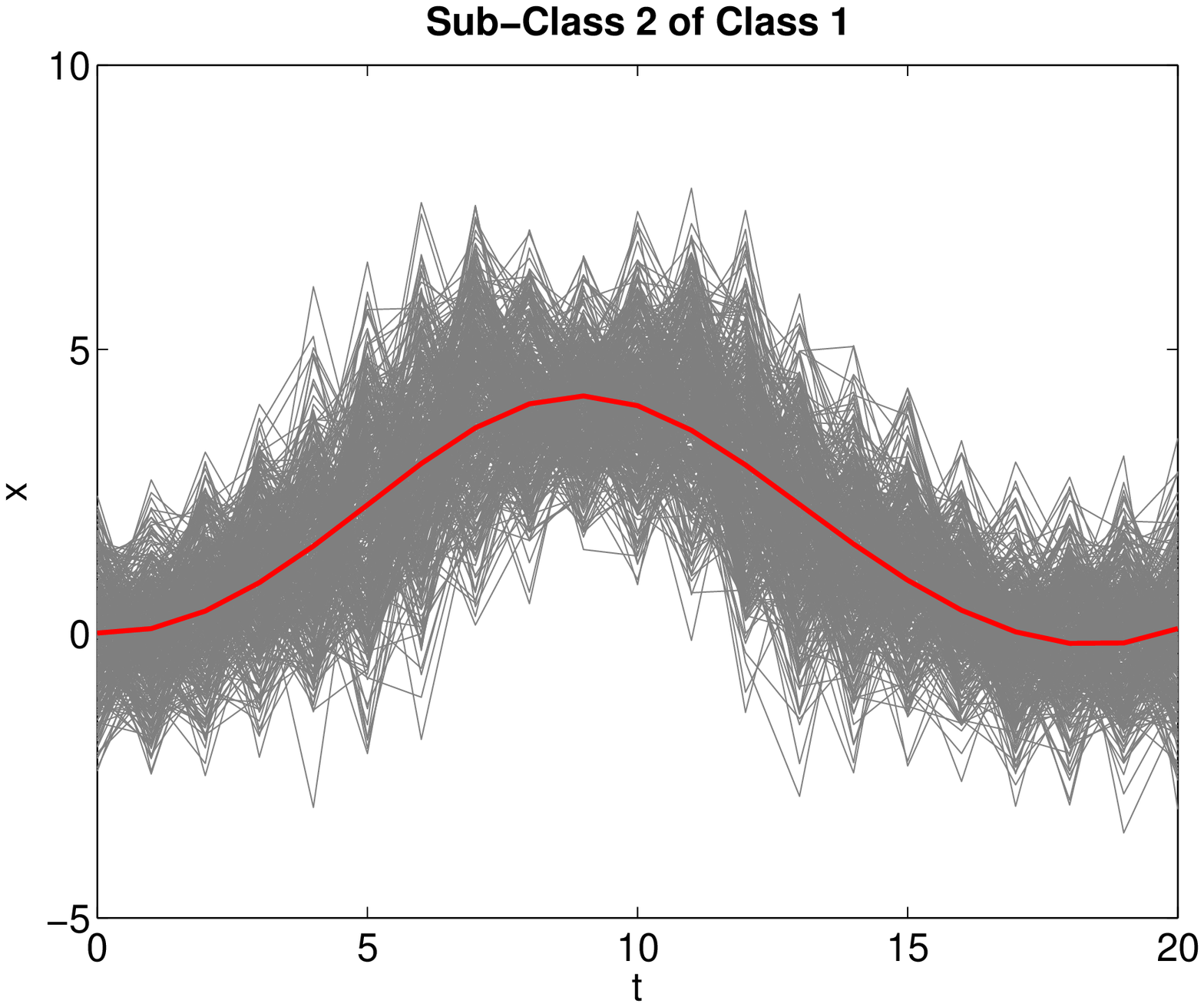} 
\includegraphics[width=4.27cm,height=3.3cm]{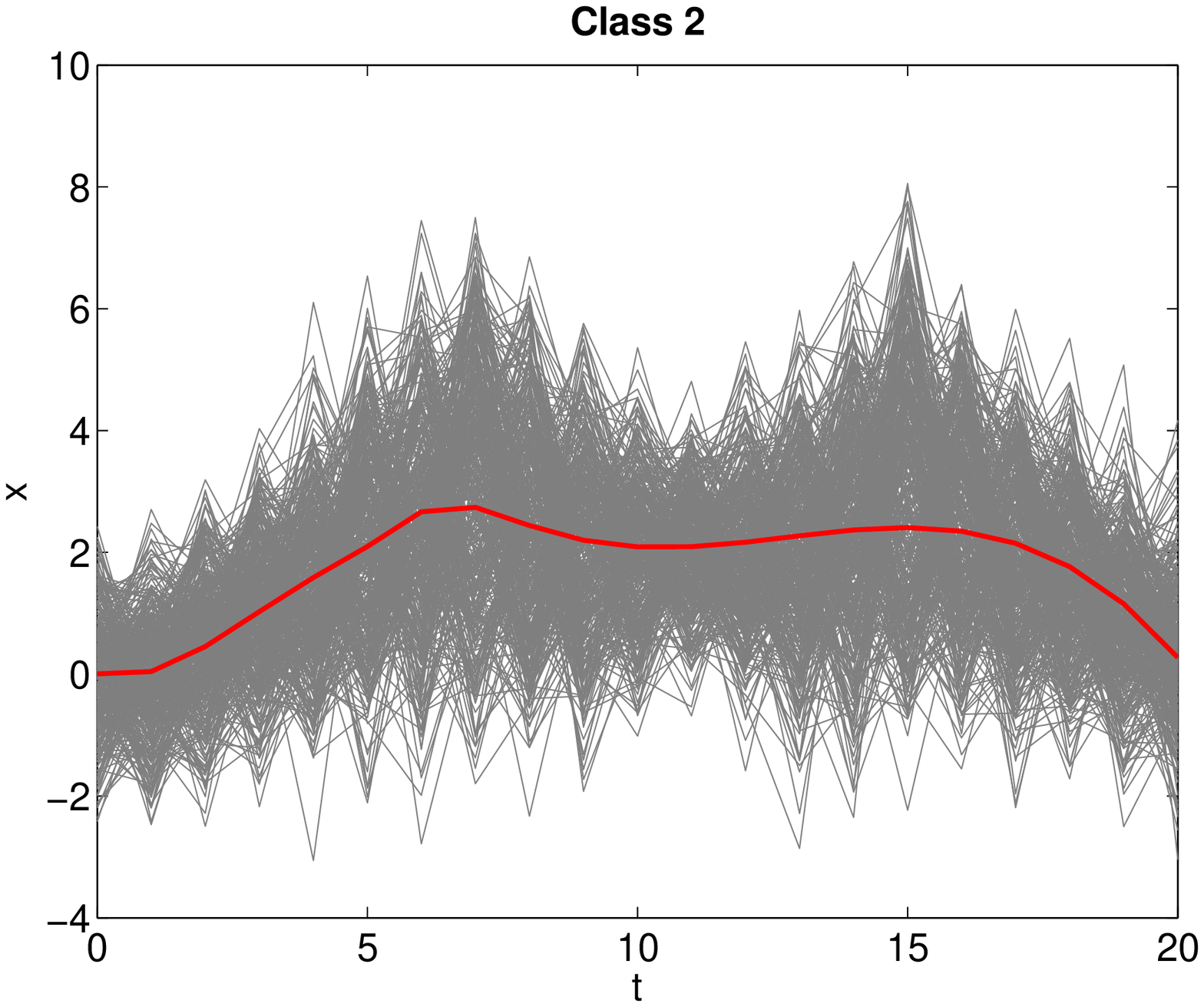}
 \caption{Modeling results for the waveform curves: (top) the waveforms (500 curves per class) where the first class is composed of two sub-classes, (middle) the waveforms and the estimated subclasses for class 1 and the corresponding mean curves for each class, and (bottom) the two subclasses of class 1 shown separately with their corresponding mean curves.}
 \label{fig: waveform curves examples and estimations}
\end{figure}

In this section, we also illustrate our proposed approach on the waveform curves. The waveform data introduced by \cite{breiman} consist of a three-class problem where each curve is generated as follows:
\begin{itemize}
\item $\bx_i(t)=uf_1(t) + (1-u)f_2(t) + \epsilon_t$ for the class 1;
\item $\bx_i(t)=uf_2(t) + (1-u)f_3(t) + \epsilon_t$ for the class 2;
\item $\bx_i(t)=uf_1(t) + (1-u)f_3(t) + \epsilon_t$ for the class 3.
\end{itemize}
where $u$ is a uniform random variable on $(0,1)$, \linebreak
$f_1(t)=\max (6-|t-11|,0)$;
$f_2(t)=f_1(t-4)$;
 $f_3(t)=f_1(t+4)$
and $\epsilon_t$ is a zero-mean Gaussian noise with unit standard deviation.
The temporal interval considered for each curve is $(0,20)$ with a constant period of sampling of 1 second. 
For the experiments considered here, in order to have a heterogeneous class, we combine both class 1 and class 2 to form a single class called class 1. Class 2 will therefore used to refer to class 3 in the previous description of the waveform data. Figure \ref{fig: waveform curves examples and estimations} (top) shows curves from each if the two classes.

Figure \ref{fig: waveform curves examples and estimations} (middle) shows the obtained modeling results for each of the two classes by applying the proposed approach. We can see that the two sub-classes for the first class are well identified. These two sub-classes (clusters) are shown separately on Figure \ref{fig: waveform curves examples and estimations} (bottom) with their corresponding mean curves. We notice that for this data set, all FMDA approaches provide  very similar results regarding both the classification and the approximation since, as it can be seen, the complexity for this example is only related to the dispersion of the first class into sub-classes, and there are no explicit regime changes; each sub-class can therefore also be accurately approximated by a polynomial or a spline function.  %

\subsection{Experiments on real data}
\label{ssec: experiments on switch curves}

\begin{figure*}[!h] 
\centering
\includegraphics[width=4.2cm,height=3.2 cm]{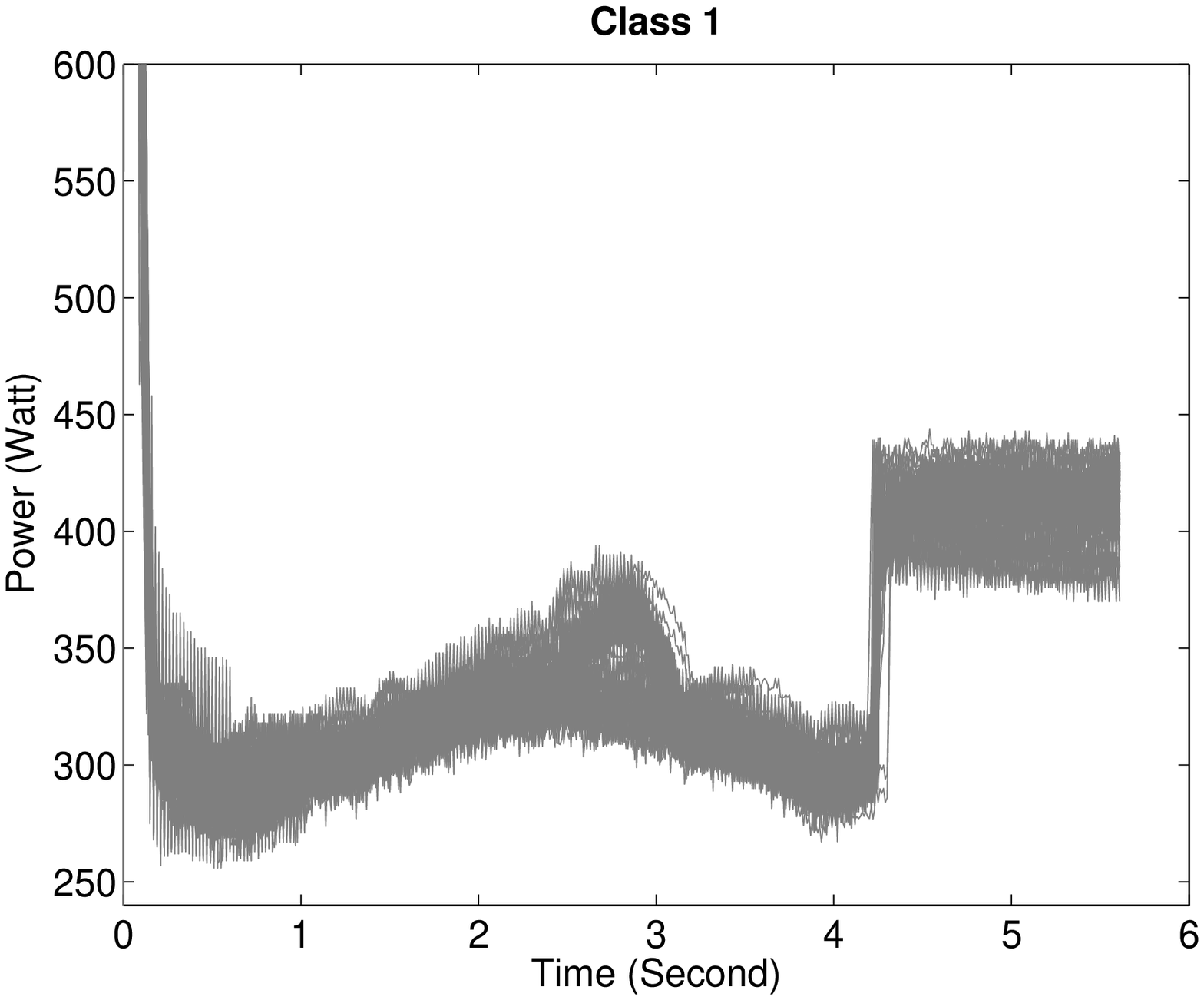}
\includegraphics[width=4.2cm,height=3.2 cm]{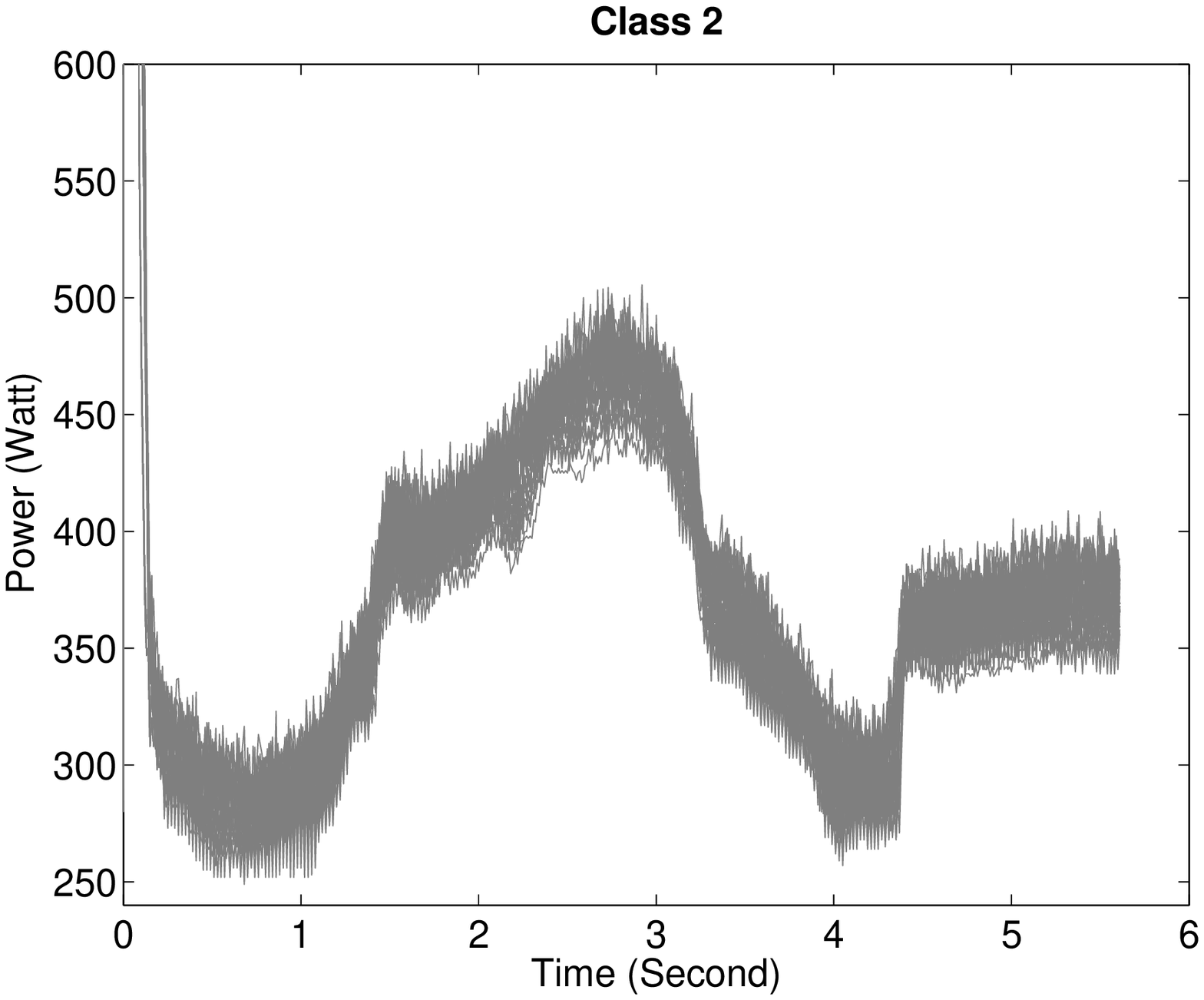} 
\caption{75 switch operation curves from the first class (left) and 45 curves from the second class (right).}
\label{fig: switch-curves-classes}
\end{figure*}

 In this section, we used a database issued from a railway diagnosis application as studied in \cite{chamroukhi_et_al_neurocomputing2010, chamroukhi_et_al_NN2009, chamroukhi_adac_2011}. 
The application context is the remote monitoring of the railway infrastructure components and more particularly the switch mechanism (also called points mechanism). The railway switch allows for guiding trains from one track to
another at a railway junction, and is driven by an electric motor.
The problem consists in accurately detecting possible defect on the the system in order to alert the maintenance services. The used data are the curves of the instantaneous electrical power consumed during the switch actuation period. Each curve consists of 564 points sampled at 100 hz in the range of (0;5.64) seconds (e.g., see Figure \ref{fig: switch-curves-classes}). 
The switch actuation consists of five successive mechanical motions of different parts of the mechanism. 
\begin{enumerate}
\item starting phase,
\item points unlocking phase, 
\item points translation phase,
\item points locking phase,
\item friction phase.
\end{enumerate}
Let us notice that the shape and the duration of each phase can vary from one situation of curves (class) to another according to the state of the system (e.g with a defect, without defect).
The used database is composed of $120$ labeled real switch operation curves. In \cite{chamroukhi_et_al_neurocomputing2010, chamroukhi_et_al_NN2009, chamroukhi_adac_2011}, the data were used to perform classification into three classes : no defect, with a minor defect and with a critical defect. In this study, we rather consider two classes where the first one is composed by the curves with no defect and with a minor defect so that the decision will be either with or without defect.  The goal is therefore to provide an accurate automatic modeling especially for class 1 which is henceforth dispersed into two sub-classes. The cardinal numbers of the classes are $n_1=75$ and  $n_2=45$ respectively. Figure \ref{fig: switch-curves-classes} shows each class of curves, where the first class is composed of two sub-classes.

\begin{figure*}[!h]
\centering
{\small \begin{tabular}{ccc}
\includegraphics[width=4.27cm,height=3 cm]{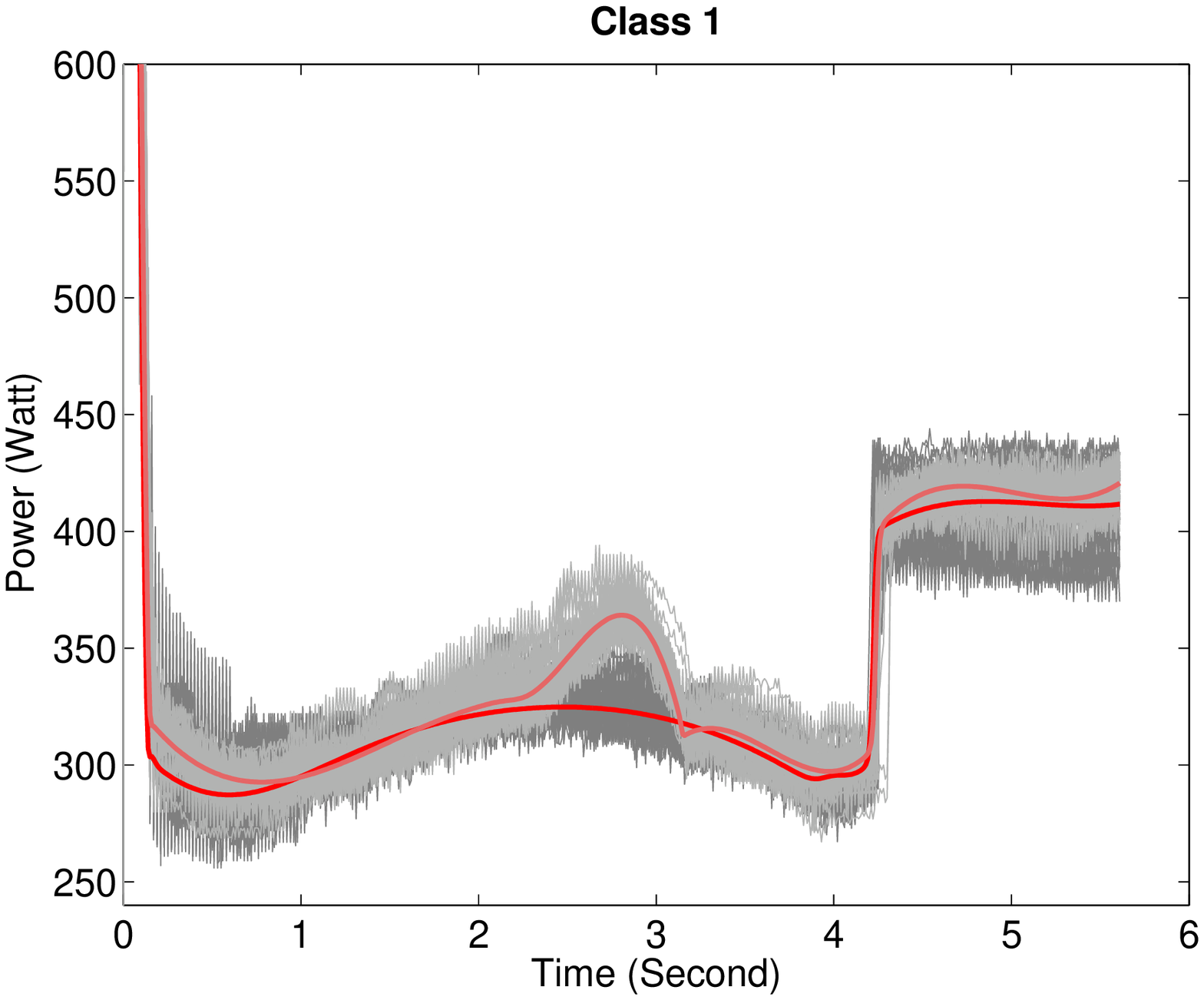}&
&
\includegraphics[width=4.27cm,height=3 cm]{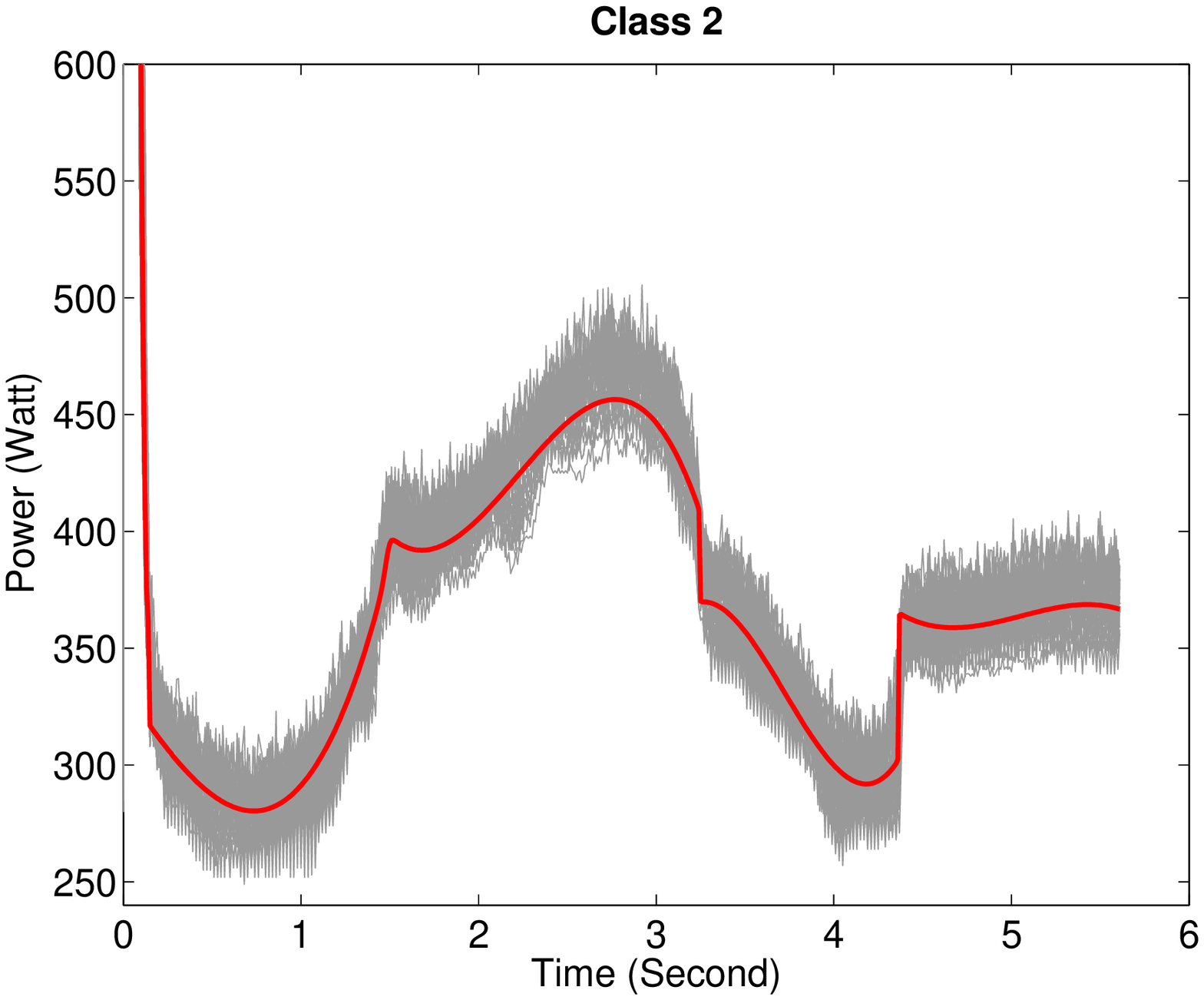}
\\
(a) & & (b) \\
\includegraphics[width=4.27cm,height=3 cm]{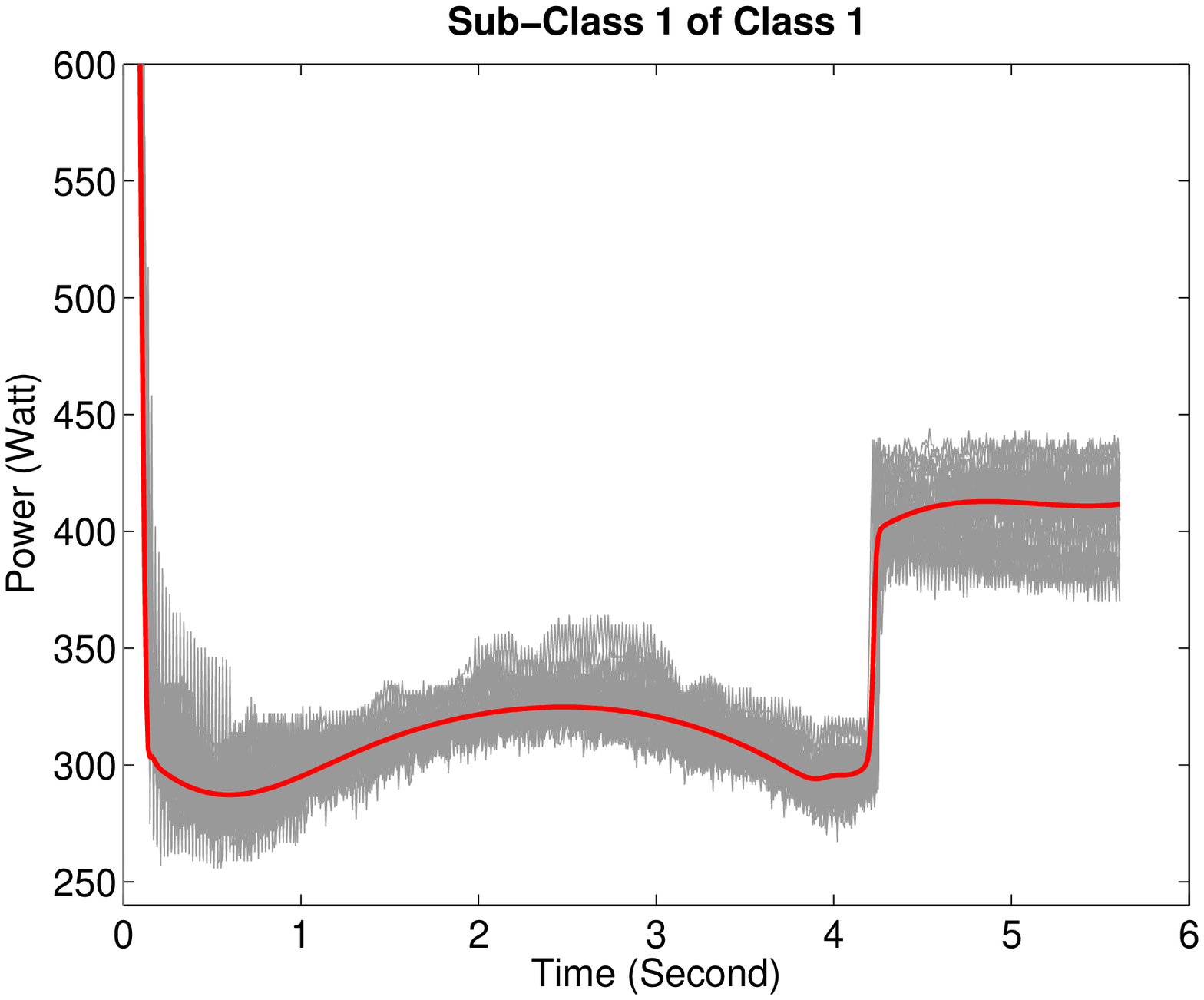}&
\includegraphics[width=4.27cm,height=3 cm]{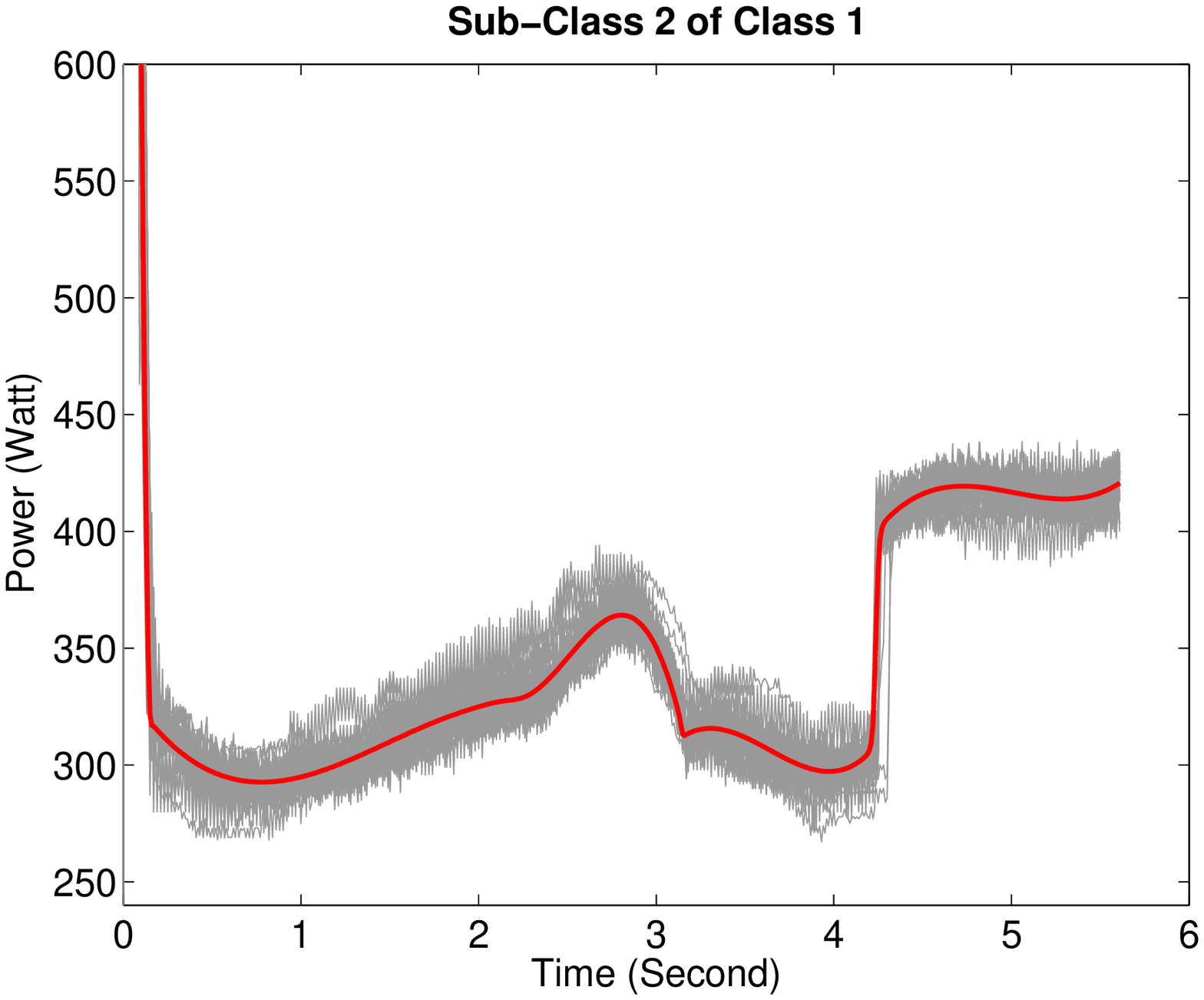}&
\includegraphics[width=4.27cm,height=3 cm]{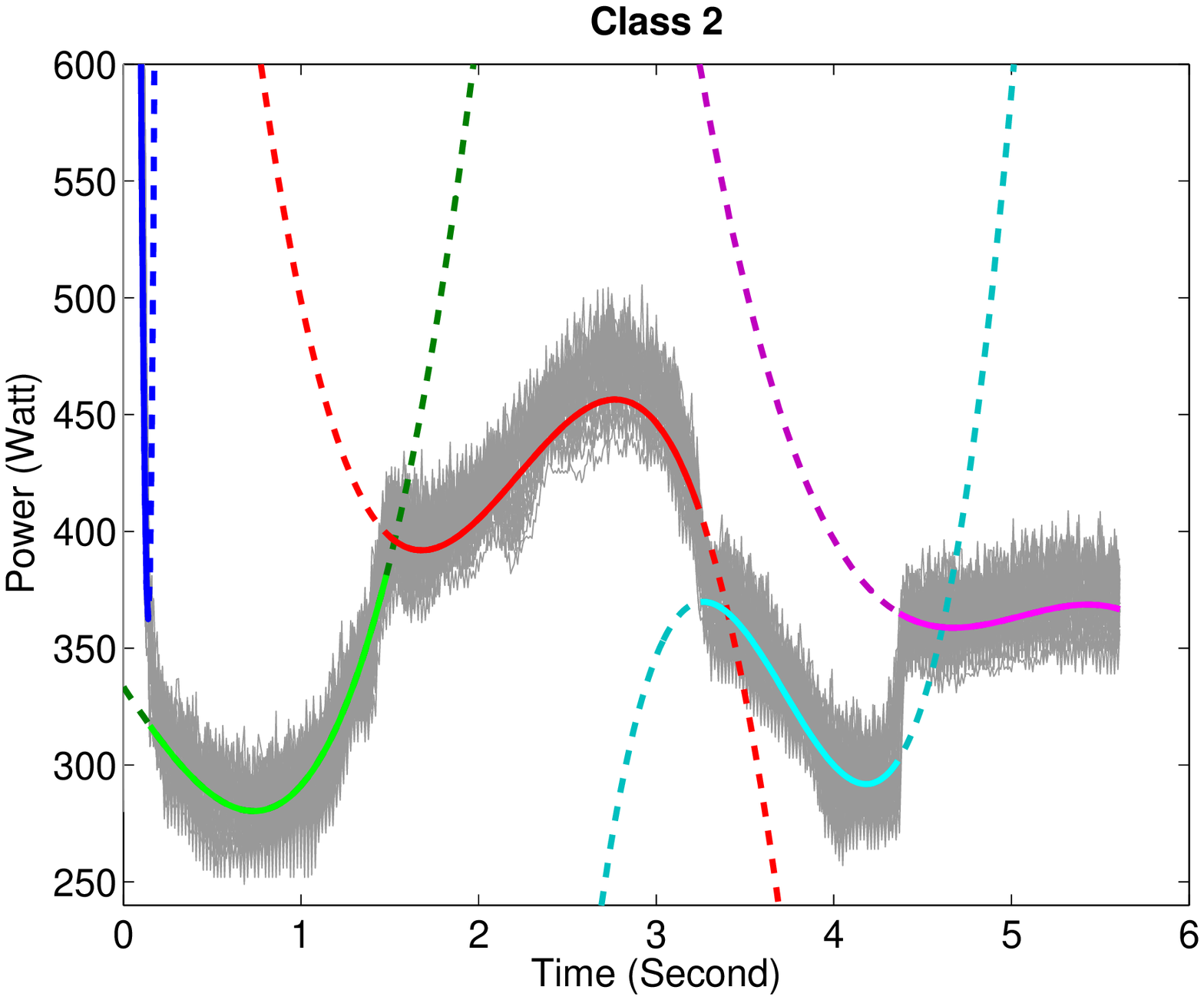}
\\
(c) & (d) & (e) \\
\includegraphics[width=4.27cm,height=3 cm]{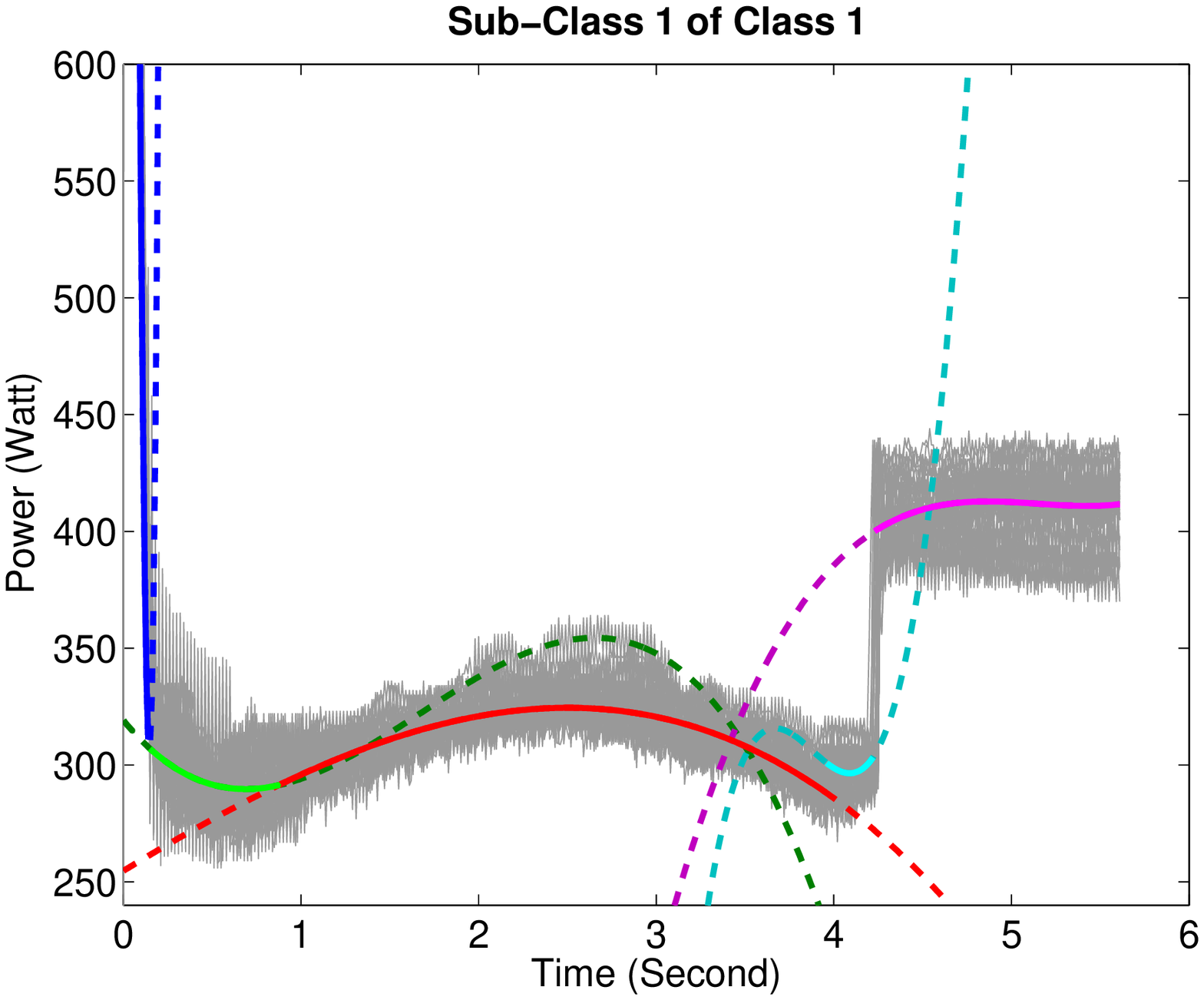}
&
\includegraphics[width=4.27cm,height=3 cm]{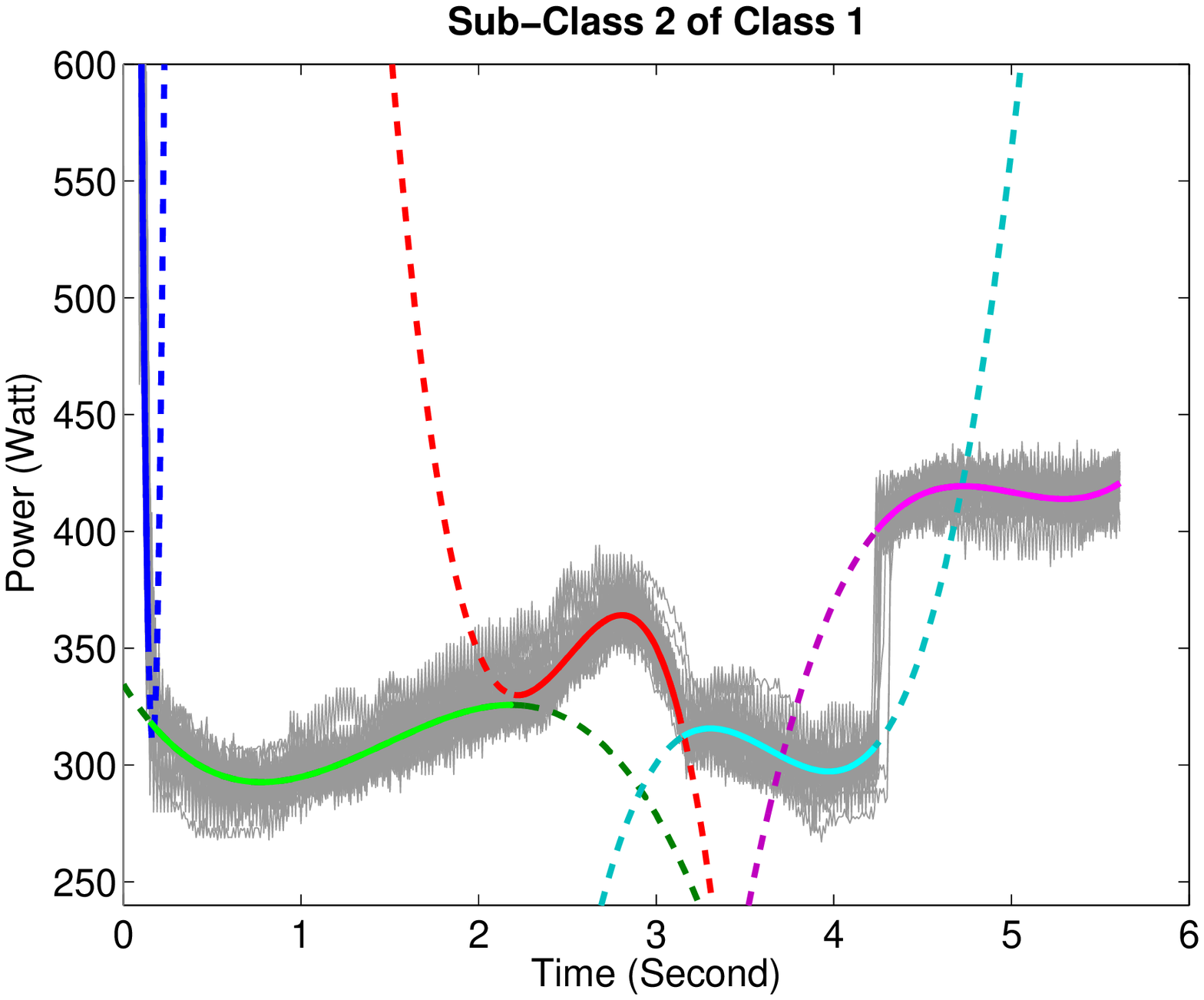}
&
\includegraphics[width=4.27cm,height=2.8 cm]{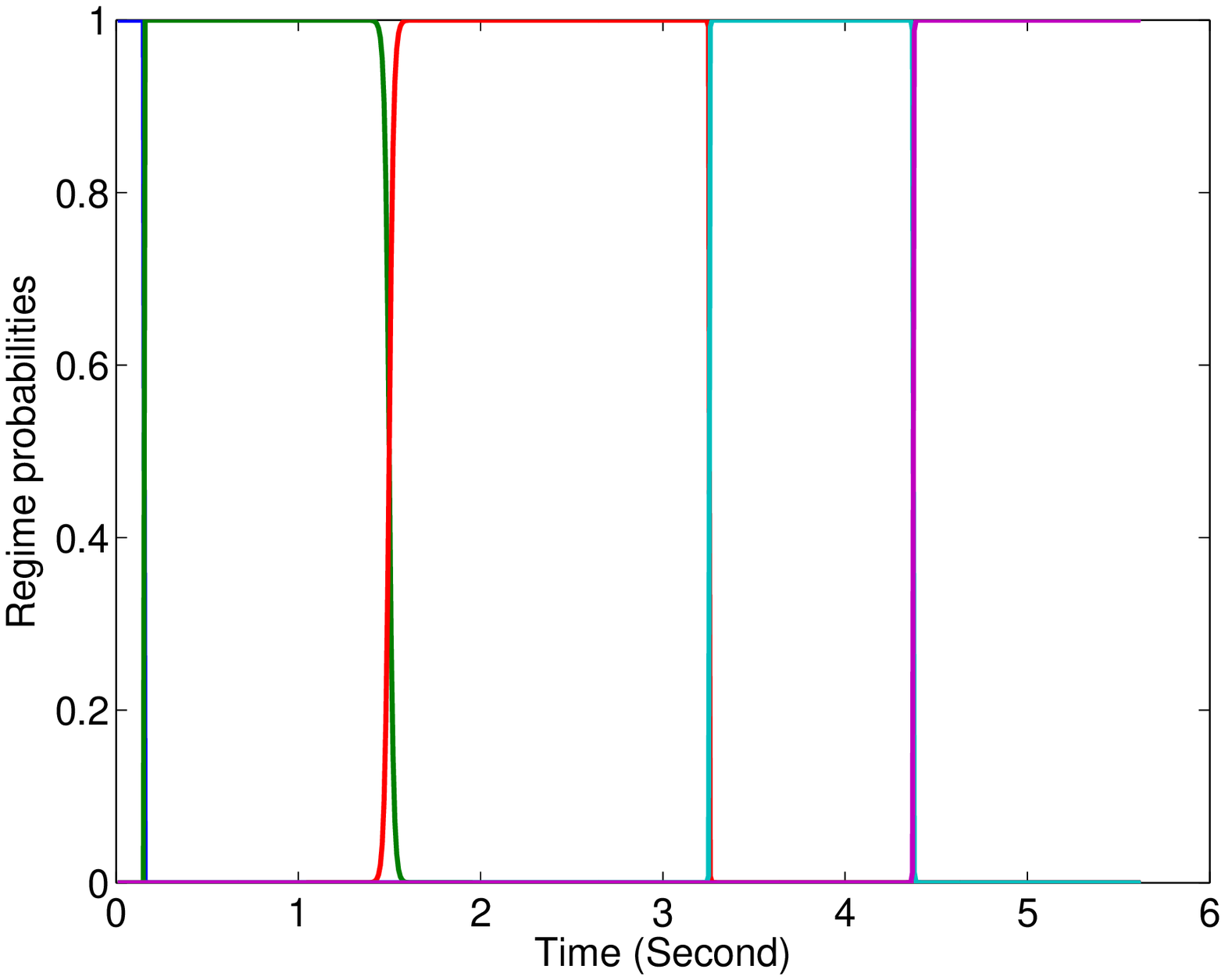}
\\
(f) & (g) & (h) \\
\includegraphics[width=4.27cm,height=2.8 cm]{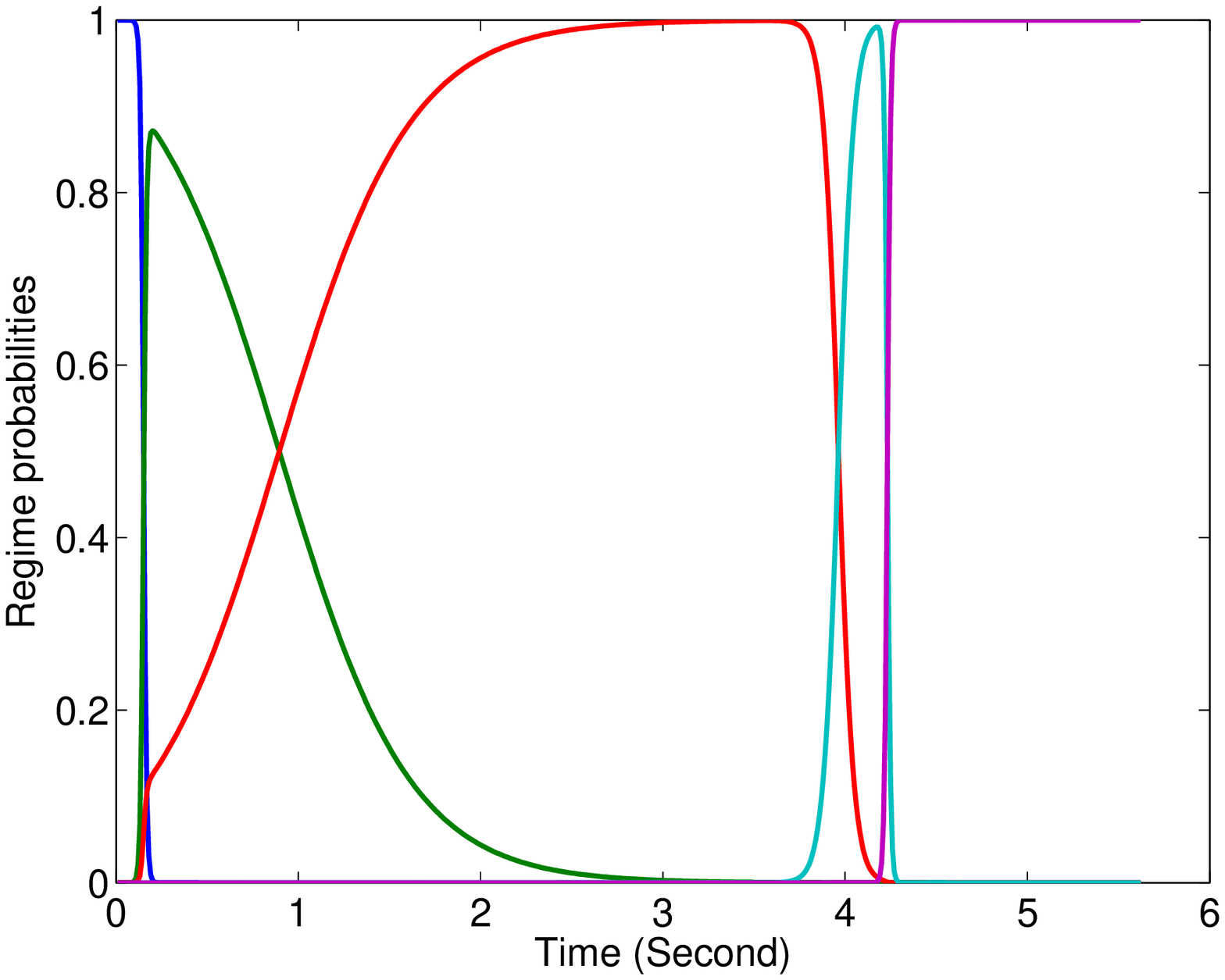}
&
\includegraphics[width=4.27cm,height=2.9 cm]{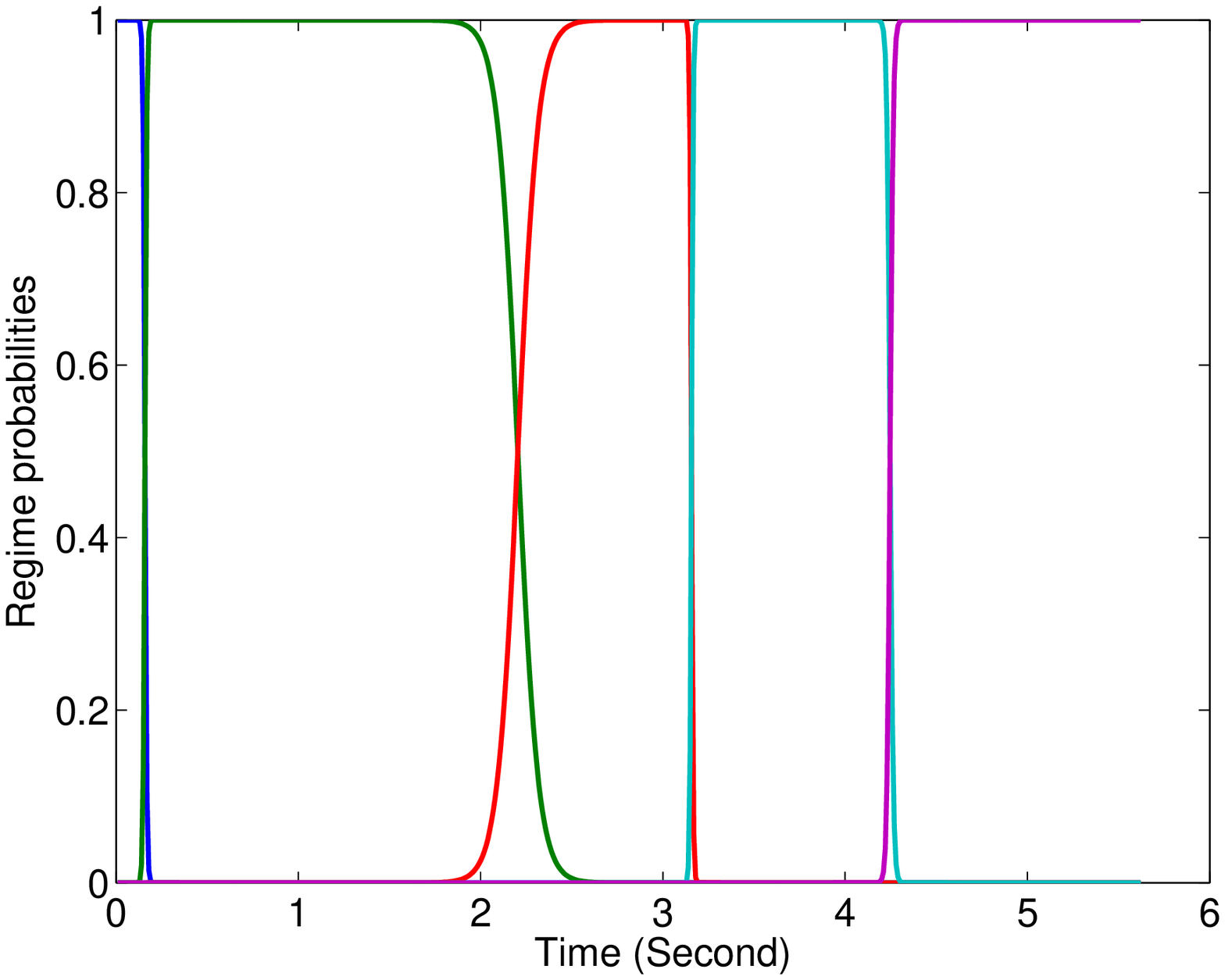}\\
(i) & (j) 
\end{tabular}}
\caption{\label{fig: switch-curves-MixRHLP results}
Results obtained with the proposed model  for the real curves.
The estimated sub-classes for class 1  and the corresponding mean curves provided by the proposed approach (a); Then, we show  separately each sub-class of class 1 with the estimated mean curve presented in a bold line (c,d), the polynomial regressors (degree $p=3$) (f,g) and the corresponding logistic proportions that govern the hidden processes (i,j). Similarly, for class 2, we show the estimated mean curve in bold line (b), the polynomial regressors (e) and the corresponding logistic proportions.} 
\end{figure*}

Figure \ref{fig: switch-curves-MixRHLP results} shows the modeling results provided by the proposed approach for each of the two classes. It shows the two sub-classes estimated for class 1 and the corresponding mean curves for the two classes. We also present the estimated polynomial regressors for each set of curves and the corresponding probabilities  of the logistic process that govern the regime changes over time. We see that the proposed method ensure both an accurate decomposition of the complex shaped class into sub-classes and at the same time, a good approximation of the underlying regimes within each homogeneous set of curves. Indeed, it can be seen that the logistic process probabilities are close to $1$ when the $\ell$th regression model seems to be the best fit for the curves and vary over time according to the smoothness degree of regime transition. 

Then, the obtained classification results, by considering the FLDA approaches and the FMDA approaches (which are more competitive) and gave the best results for simulations, are given in Table \ref{table: results for switch curves}. 
{\small \begin{table}[!h]
\centering
\begin{tabular}{|l|c|c|} 
\hline
Approach &  Classification error rate (\%) & Intra-class inertia\\ 
\hline 
\hline 
FLDA-PR   & 11.5   &  $10.7350 \times 10^9$ \\
FLDA-SR  & 9.53 &  $9.4503 \times 10^9$ \\
FLDA-RHLP & 8.62 &  $8.7633\times 10^9$ \\
\hline 
FMDA-PRM 	& 9.02 &    $7.9450 \times 10^9$\\
FMDA-SRM     & 8.50 & $5.8312 \times 10^9$ \\
{\bf FMDA-MixRHLP } & {\bf 6.25} & ${\bf 3.2012 \times 10^9}$ \\
\hline
\end{tabular}
\caption{\label{table: results for switch curves}
Obtained results for the real curves of switch operations.}
\end{table}}

We can see that, although the classification results are similar for the FMDA approaches, the difference in terms of curves modeling (approximation) is significant, for which the proposed FMDA-MixRHLP approach clearly outperforms the alternative ones. This is attributed to the fact that the use of polynomial regression mixtures for FMDA-PRM or spline regression mixtures (FMDA-SRM) does not fit at best the regime changes compared to the proposed model. However, even the proposed approach provides the better results, we note that we have many parameters to estimate as summarized by table one for the complex class (class 1). 
 {\small \begin{table}[!h]
\centering
\begin{tabular}{|l|c|c|} 
\hline
Model &  Number of parameters for class 1\\ 
\hline 
\hline 
FLDA-PR   & 5\\
FLDA-SR  & 15 \\
FLDA-RHLP & 33\\
\hline 
FMDA-PRM 	& 11\\
FMDA-SRM     & 31\\
{\bf FMDA-MixRHLP } & 67\\
\hline
\end{tabular}
\caption{\label{table: }Number of free parameters for each of the used models for class 1.}
\end{table}}On the other hand, we note that for these values of $(K,L,p)$ provided by the experts, there is no over-fitting.

We also note that, for this real data, in terms of required computational effort to train each of the compared methods, the FLDA approaches are faster than the FMDA ones. In FLDA, both the polynomial regression and the spline regression approaches are analytic and does not require an numerical optimization scheme. The FLDA-RHLP is based on an EM algorithm which, while therefore performs in an iterative way, the learning scheme is quite fast and is in mean around one minute for the described real data, and outperforms the alternative piecewise regression using dynamic programming and significantly improves the results. Detailed comparisons have been given in \cite{chamroukhi_et_al_neurocomputing2010}, namely in terms of computational time. 
On the other hand, the alternative FMDA approaches, that is the regression mixture and the spline regression mixture-based approaches still more fast and their EM algorithm requires only few seconds to converge. However, these approaches are clearly not adapted for the regime changes problem; to do that, one needs to built a piecewise regression-based model which requires dynamic programming and therefore a dramatical computational cost especially for large curves, and still only adapted to abrupt regime changes. 
As stated in section \ref{ssec: model selection method}, the training procedure for the proposed approach is not dramatically time consuming, the training for the data of class 1 (which is the more complex class), requires a mean computational time of 2.98 minutes on a Matlab software using a standard laptop CPU of 2Ghz.


For model selection for this real dataset, we notice that the number of regimes is fixed by the experts ($L=5$) and equals the number of electromechanical phases of the switch operation \citep{chamroukhi_et_al_NN2009, chamroukhi_et_al_neurocomputing2010}. The number of sub-classes for class 1 is $K=2$ as we have no-defect sub-class and a sub-class corresponding to curves with a minor defect. The polynomial degree which is well adapted to the regime shape for the curves is $p=3$ (this was a preliminary choice made
in conjunction with the expert \citep{chamroukhi_et_al_NN2009, chamroukhi_et_al_neurocomputing2010} and a model selection  procedure in \cite{chamroukhi_adac_2011} have confirmed this choice.

\section{Conclusion}
\label{sec: conclusion}
In this paper, we presented a new mixture model-based approach for functional data classification. The discrimination approach includes an unsupervised task that consists in clustering dispersed classes into sub-classes and determining the underlying unknown regimes for each sub-class.
The proposed functional discriminant analysis approach uses a specific mixture of hidden process regression model for each class, which is particularly adapted for modeling complex-shaped classes of curves presenting regime changes. 
The parameters of each class are estimated  in an unsupervised way by a dedicated EM algorithm and a model selection procedure is presented.  The experimental results on simulated data and real data and comparisons to alternative approaches demonstrate the effectiveness of the proposed approach. 
In a first time, a future work will mainly concern learning the MixRHLP model of each class into by maximzing a classification likelihood criterion, in which we will mainly be interested into classification, rather than maximizing a likelihood criterion as in this approach where we mainly focus on model estimation. This will rely on the Classification EM (CEM) algorithm \citep{celeuxetgovaert92_CEM}.  
Another future perspective is to build a fully Bayesian model for functional data to explicitly incorporate some prior knowledge on the data structure and to better control the model complexity. 

\bibliographystyle{elsarticle-num}
\bibliography{references}
\end{document}